\documentclass[usenatbib]{mnras}

\usepackage[T1]{fontenc}
\usepackage{ae,aecompl}

\usepackage{graphicx}	% Including figure files
\usepackage{amsmath}	% Advanced maths commands
\usepackage{amssymb}	% Extra maths symbols
\usepackage{mathrsfs}
\usepackage{enumitem}
\usepackage[normalem]{ulem} % Editing
\usepackage{multirow} % Merging table cells
\usepackage{xcolor} % Highlighting
\usepackage{lipsum} % Lorem Ipsum
\usepackage{hyperref} % Eprint and doi bibtex fields shall be hyperlinks
\usepackage{mathtools} % For \coloneqq command etc.
\usepackage{subcaption}
\usepackage{array} 
\usepackage{lipsum}
\usepackage{txfonts}
\usepackage{float}
\usepackage{physics}
\usepackage[shortcuts]{extdash}
\usepackage{lipsum}  
\usepackage{dblfloatfix}
\usepackage{siunitx}
\usepackage{booktabs}
\usepackage{amstext} % for \text macro
\usepackage{array}   % for \newcolumntype macro
\newcolumntype{L}[1]{>{\centering\arraybackslash$}p{#1}<{$}}

\newcommand{\TNG}{\mbox{\scshape{IllustrisTNG}}\normalfont}
\newcommand{\VELA}{\mbox{\scshape{VELA}}\normalfont}
\newcommand{\IBox}{\mbox{\scshape{TNG100}}\normalfont}
\newcommand{\SBox}{\mbox{\scshape{TNG50}}\normalfont}
\newcommand{\BBox}{\mbox{\scshape{TNG100+50}}\normalfont}
\newcommand{\FL}{\mbox{\scshape{FirstLight}}\normalfont}
\newcommand{\FLBBox}{\mbox{\scshape{FirstLight10+20}}\normalfont}

\DeclareRobustCommand{\VAN}[3]{#2}
\let\VANthebibliography\thebibliography
\def\thebibliography{\DeclareRobustCommand{\VAN}[3]{##3}\VANthebibliography}

\makeatletter
\def\footnoterule{\kern-3\p@
  \hrule \@width 2in \kern 2.6\p@} % the \hrule is .4pt high
\makeatother

\title[Mini-Quenching of Galaxies at $z=4-8$]{Mini-Quenching of $z=4-8$ Galaxies by Bursty Star Formation}

\author[T. Dome et al.]
{\parbox[t]{\textwidth}
{Tibor Dome$^{1,2}$\thanks{E-mail: td448@cam.ac.uk}, 
Sandro Tacchella$^{2,3}$, 
Anastasia Fialkov$^{1,2}$,
Daniel Ceverino$^{4,5}$,
Avishai Dekel$^{6,7}$,
Omri Ginzburg$^{6}$, 
Sharon Lapiner$^{6}$,
Tobias J. Looser$^{2,3}$
}
\\ \\
$^{1}$Institute of Astronomy, University of Cambridge, Madingley Road, Cambridge, CB3 0HA, UK\\
$^{2}$Kavli Institute for Cosmology, Madingley Road, Cambridge, CB3 0HA, UK\\
$^{3}$Cavendish Laboratory, University of Cambridge, 19 JJ Thomson Avenue, Cambridge, CB3 OHE, UK\\
$^{4}$Departamento de Fisica Teorica, Modulo 8, Facultad de Ciencias, Universidad Autonoma de Madrid, E-28049 Madrid, Spain\\
$^{5}$CIAFF, Facultad de Ciencias, Universidad Autonoma de Madrid, E-28049 Madrid, Spain\\
$^{6}$Center for Astrophysics and Planetary Science, Racah Institute of Physics, The Hebrew University, Jerusalem 91904, Israel\\
$^{7}$SCIPP, University of California, Santa Cruz, CA 95064, USA
}

\date{Accepted XXX. Received YYY; in original form ZZZ}

\pubyear{2023}

\begin{document}
\label{firstpage}
\pagerange{\pageref{firstpage}--\pageref{lastpage}}
\maketitle

\begin{abstract}
The recent reported discovery of a low-mass $z=5.2$ and an intermediate-mass $z=7.3$ quenched galaxy with JWST/NIRSpec is the first evidence of halted star formation above $z\approx 5$. Here we show how bursty star formation at $z=4-8$ gives rise to temporarily quenched, or mini-quenched galaxies in the mass range $M_{\star} = 10^7-10^9 \ M_{\odot}$ using four models of galaxy formation: the periodic box simulation \TNG, the zoom-in simulations \VELA \ and \FL \ and an empirical halo model. The main causes for mini-quenching are stellar feedback, lack of gas accretion onto galaxies and galaxy-galaxy interactions. The abundance of (mini-)quenched galaxies agrees across the models: the population first appears below $z\approx 8$, after which their proportion increases with cosmic time, from $\sim 0.5-1.0$\% at $z=7$ to $\sim 2-4$\% at $z=4$, corresponding to comoving number densities of $\sim 10^{-5}$ Mpc$^{-3}$ and $\sim 10^{-3}$ Mpc$^{-3}$, respectively. These numbers are consistent with star formation rate duty cycles inferred for \VELA \ and \FL \ galaxies. Their star formation histories (SFHs) suggest that mini-quenching at $z=4-8$ is short-lived with a duration of $\sim 20-40$ Myr, which is close to the free-fall timescale of the inner halo. However, mock spectral energy distributions of mini-quenched galaxies in \TNG \ and \VELA \ do not match JADES-GS-z7-01-QU photometry, unless their SFHs are artificially altered to be more bursty on timescales of $\sim 40$ Myr. Studying mini-quenched galaxies might aid in calibrating sub-grid models governing galaxy formation, as these may not generate sufficient burstiness at high redshift to explain the SFH inferred for JADES-GS-z7-01-QU.
\end{abstract}


%%%%%%%%%%%%%%%%%%%%%%%%%%%%%%%%%%%%%%%%%%%%%%%%%%

%%%%%%%%%%%%%%%%% BODY OF PAPER %%%%%%%%%%%%%%%%%%

\section{Introduction}
\label{s_intro}

The tight relation between stellar mass and star formation rate (SFR) of galaxies, known as the star-forming main sequence (MS), persists out to high redshift \citep{Speagle_2014, Silva_2023, Popesso_2023}, even if the shape, scatter and normalisation have yet to be accurately determined above redshift $z\approx5$. For star-forming galaxies, the `regulator' or `bathtub' model posits that the SFR in a galaxy is controlled by the self-regulation of gas inflow, gas outflow, and gas consumption through star formation, providing a framework for understanding continuous star formation over long timescales on the MS \citep{Bouche_2010, Dekel_2013_b, Lilly_2013, Tacchella_2016}. Massive galaxies at and above the knee of the galaxy stellar mass function, then reduce their star-formation activity and leave the MS. Those galaxies are typically called `quiescent' (or `quenched') and increase  significantly in abundance from redshift $z\approx2$ to today \citep{Faber_2007, Peng_2010}.\par

A plethora of mechanisms has been proposed to explain the quenching of galaxies, each acting on different timescales and classes of galaxies. To match the observed galaxy number densities, \textit{internal mechanisms} \citep{Croton_2006, Merlin_2012, Sherman_2020, Zinger_2020, Tacchella_2022} are invoked at both the low-mass and high-mass end of the galaxy stellar mass function. Its suppression below the knee \cite[$M_{\star} \sim 10^{10.5} \ M_{\odot}$ at $z\approx 4$,][]{Grazian_2015} is typically accomplished via stellar feedback and above the knee via feedback from active galactic nuclei \citep[AGN,][]{Curtis_2016, Henden_2018, Nelson_2021}, both of which expel gas and heat the circumgalactic medium. Some of these internal mechanisms, such as bulge formation, could lead to \textit{morphological quenching} \citep{Martig_2009, Gensior_2020, Lu_2021, Shin_2022}. Since star formation preferentially occurs in gravitationally unstable gas discs, the stabilization thereof by the presence of the bulge (its growth and evolution regulated by AGN feedback) effectively reduces the star-formation activity. Especially at low redshift of $z<2$ where galaxies can fall into dense clusters, \textit{environmental quenching} effects \citep{Dekel_2006, Peng_2012, Ji_2018, Contini_2020, Whitaker_2021, Williams_2021} such as the loss or removal of gas from a galaxy due to ram pressure, tidal interactions and virial shock heating (all three processes on timescales of $100$ Myr $-2$ Gyr) are believed to play a major role in regulating SFRs of low-mass $(M_{\star} < 10^{10} M_{\odot})$ satellite galaxies.\par

Since density contrasts between (proto-)clusters and the field are relatively small at high redshift of $z>2$ \citep{Overzier_2016}, environmental quenching is less efficient. Nonetheless, quiescent galaxies have been detected at much higher redshift with their number density predicted to correlate inversely with redshift, down to $\sim 10^{-5.5}$ Mpc$^{-3}$ \citep[CANDELS,][]{Merlin_2019} or $\sim 10^{-4.5}$ Mpc$^{-3}$ \citep[JWST CEERS,][]{Carnall_2023_b} at $z\approx 5$. Most of the early quiescent galaxies that have been identified to date \citep{Glazebrook_2017, Carnall_2020, Forrest_2020, Valentino_2020, Santini_2021, Nanayakkara_2022, Long_2023} are massive ($M_{\star} > 10^{10} \ M_{\odot}$), possibly due to observational limitations. Specifically, the observability of galaxies near a survey's limiting flux (typically bursty low-mass and/or high-redshift galaxies) can be highly time-dependent due to the SFR variability \citep{Sun_2023}.\par 

Recently, \cite{Looser_2023} reported the discovery of a quiescent galaxy ($\log_{10}(\mathrm{SFR} [M_{\odot}/\mathrm{yr}]) = -2.6^{+1.5}_{-2.7}$) at a reionization-era redshift of $z=7.3$ by analysing the H$\beta$ and [OIII]$\lambda5008$ emission-line fluxes as observed with JWST/NIRSpec. The mass of this quiescent galaxy is $M_{\star} = 5.0^{+1.3}_{-1.0} \times 10^8 \ M_{\odot}$, which implies a specific SFR (sSFR) of $\log_{10}(\mathrm{sSFR} [\mathrm{yr}^{-1}]) = -11.3^{+1.6}_{-2.8}$. The second-highest redshift quiescent / post-starburst galaxy reported to date is at $z=5.2$ \citep{Strait_2023} with an even lower stellar mass (for the main bulge) of $M_{\star} = 4.3^{+0.9}_{-0.8} \times 10^7 \ M_{\odot}$ and SFR$(\mathrm{H}\alpha) = 0.14^{+0.17}_{-0.12} \ M_{\odot}/\mathrm{yr}$. This implies $\log_{10}(\mathrm{sSFR} [\mathrm{yr}^{-1}]) = -8.5^{+0.4}_{-0.9}$. Observationally, it is difficult to assess whether these two galaxies will be permanently quenched or whether they will rejuvenate and return to the MS. Also note the different timescales probed by the emission lines. Short timescales are traced by H$\alpha$ and H$\beta$, intermediate ones ($\sim 10$ Myr) by UV continuum tracers while longer ones typically by near-infrared to far-infrared (NIR-FIR) indicators \citep{Speagle_2014, Katsianis_2020, Caplar_2019, Tacchella_2022_b}.\par 

Can the population of low-mass high-redshift quiescent galaxies be explained with simple models of bursty star formation? At high redshift and lower stellar masses, an equilibrium between stellar feedback and gravity cannot be sustained, leading to bursty star formation \citep{Alcazar_2017, Faucher_2018}. Therefore, although the galaxy population might still follow the MS at high redshift, their trajectory about the MS could be dominated by short-term bursts of star formation triggered by various processes such as the merging of galaxies or the inflow of gas \citep{Puebla_2016, Tacchella_2016, Tacchella_2020}. These bursts are separated by periods of relative quiescence in which galaxies fall off the star-forming MS. It is therefore of great interest to infer SFRs that probe different timescales in order to observationally constrain the burstiness of low-mass and high redshift galaxies \citep{Weisz_2012, Emami_2019, Faisst_2019}.\par

The discoveries by \cite{Looser_2023} and \cite{Strait_2023} motivated us to investigate the properties of high-redshift quiescent galaxies with low stellar masses ($M_{\star} = 10^7-10^9 \ M_{\odot}$) using theoretical models. Specifically, we use four galaxy formation models: the periodic box simulation \TNG \ \citep{Pillepich_2018, Pillepich_2019, Nelson_2019},  the zoom-in simulation \VELA \ \citep{Ceverino_2014, Zolotov_2015} and an empirical halo model (EHM; \citealt{Tacchella_2018}). These four models allow us to study this galaxy population from complementary viewpoints since they provide access to different dynamic ranges and timescales. The combination of those four models allows us to cross-check the consistency of predictions on the theory side. All four models produce quiescent galaxies with $M_{\star} = 10^7-10^9 \ M_{\odot}$ at $z>4$. However, these galaxies are only temporarily quiescent and rejuvenate back onto the MS. In order to differentiate this process from permanent\footnote{Rejuvenation of massive galaxies is rare \citep[e.g.,][]{Chauke_2019, Tacchella_2022}.} star formation quenching at higher stellar masses ($M_{\star} > 10^{10} \ M_{\odot}$), we refer to this temporary quenching as \textit{mini-quenching}. We show in this work that the main causes for mini-quenching are stellar feedback, lack of gas accretion onto galaxies, mergers and tidal interactions. The abundance of these mini-quenched galaxies (MQGs) is in good agreement between the four galaxy formation models. However, we find that the level of burstiness inferred from the four models on short timescales $\sim 40$ Myr is lower than what is observed.\par

\begin{figure*}
\includegraphics[width=\textwidth]{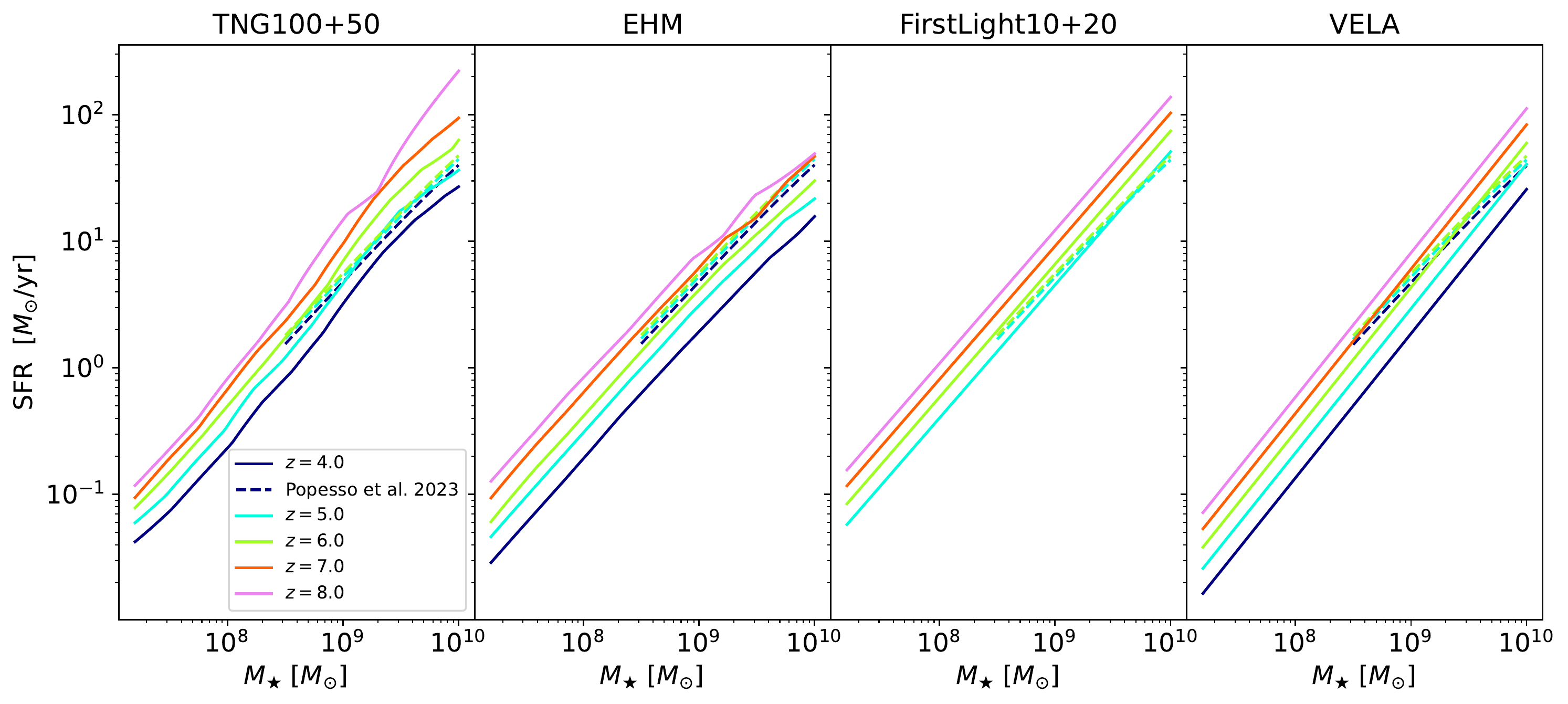}
\caption{The star-forming MS in the four galaxy formation models investigated in this work across redshifts $z=4-8$: the combined simulation study \BBox \ (first panel), EHM (second panel), \FLBBox \ (third panel) and \VELA \ (fourth panel). SFRs are averaged over $10$ Myr. A compilation of observational studies by \protect\cite{Popesso_2023} is given in the form of a common parametrization (dashed curves) below $z\leq 6$ for $M_{\star} \geq 10^{8.5} \ M_{\odot}$. The MS ridges in the four models are consistent to within $\sim 30$\%, with \BBox \ and \FLBBox \ predictions closest to observational constraints.}
\label{f_ms}
\end{figure*}

The organisation of the paper is as follows: In Secs. \ref{ss_tng_sims}, \ref{ss_em}, \ref{ss_fl} and \ref{ss_vela_sims}, we describe the four galaxy formation models: \TNG, EHM, \FL \ and \VELA. While the periodic box simulation \TNG \ provides a statistical sample for estimating abundances, \FL \ and \VELA \ zoom-ins of galaxies have more resolved star-formation histories (SFHs). We strive to treat all four models on an equal footing and present their differences in Sec. \ref{ss_model_diffs}. Details on how we calculate stellar masses and star-formation rates can be found in Secs. \ref{ss_av_timescale}, \ref{ss_aperture} and \ref{ss_ms}. Methods to generate mock spectral energy distributions (SEDs) are described in Sec. \ref{ss_seds}. In Sec. \ref{s_what_is_miniquenching}, we highlight that the properties, abundances and timescales of mini-quenching events at high redshift are roughly consistent across all four models, which we show by providing quantitative estimates. We compare simulated SEDs to JADES-GS-z7-01-QU photometry in Sec. \ref{s_sed_jades}. We discuss our main findings in \mbox{Sec. \ref{s_conclusions}}.

\section{Theoretical Models and Post-Processing}
\label{s_num_methods}

\subsection{IllustrisTNG Simulations}
\label{ss_tng_sims}
In order to study a representative sample of galaxies, we use both \mbox{\scshape{IllustrisTNG50}} \normalfont and \mbox{\scshape{IllustrisTNG100}}. \normalfont These simulations were performed with the state-of-the-art code \mbox{\scshape{Arepo}} \normalfont described by \cite{Arepo_2010} and \cite{Weinberger_2020}. The hydrodynamical equations are solved on a moving Voronoi mesh using a finite volume method. Various astrophysical processes such as metal-line cooling, star formation and feedback remain unresolved in \TNG \ and are approximated by subgrid models \citep{Pillepich_2017}. Gas above a density threshold of $n_{\text{H}} \sim 0.1$ cm${}^{-3}$ forms stars stochastically following the empirical Kennicutt-Schmidt relation and assuming a Chabrier \citep{Chabrier_2003} initial mass function (IMF), see Table \ref{t_sims}.\par 

The dark matter particle mass resolution is $m_{\text{DM}} = 7.5 \times 10^6 \ M_{\odot}$ in \IBox \ and $m_{\text{DM}} = 4.5 \times 10^5 \ M_{\odot}$ in \SBox. Dark matter haloes are identified using the friends-of-friends ({\fontfamily{cmtt}\selectfont FoF}) algorithm with a standard linking length of $b = 0.2 \times \text{(mean inter-particle separation)}$ \citep{Springel_2001_2}. Within each FoF halo, sub-haloes identified by the SUBFIND \citep{Springel_2001} algorithm are made up of all the resolution elements (gas, stars, dark matter, and black holes) which are gravitationally bound to the subhalo. In our framework, galaxies are subhaloes with at least $\sim 200$ stellar particles, i.e. $M_{\star,\text{min}} = 2\times 10^7 \ M_{\odot}$ for \SBox \ and $M_{\star,\text{min}} = 2\times 10^8 \ M_{\odot}$ for \IBox. The sample includes both central and satellite galaxies. We also perform a combined simulation study of \BBox, with appropriate weights when calculating comoving number densities of MQGs. We disregard \mbox{\scshape{TNG300}} \normalfont since sufficiently resolved galaxies ($M_{\star} > 2 \times 10^9 \ M_{\odot}$) are not expected to exhibit mini-quenching events in the simulation (see Sect. \ref{ss_drivers}), even though some do exhibit AGN-induced quiescence by $z\approx 4.2$ \citep{Hartley_2023}.

\subsection{Empirical Halo Model}
\label{ss_em}
Empirical models are galaxy formation models whose physical prescription is largely motivated and calibrated based on observations. The workings for semi-analytical and empirical models have become increasingly similar, though the former solves physical equations, while the later focuses on effective prescriptions. Empirical models are now successful in describing the galaxy population over a wide range of redshifts \citep{Puebla_2016, Behroozi_2019, Tacchella_2013, Tacchella_2018}. We will focus on the empirical model introduced in \cite{Tacchella_2018}, based on halo merger trees extracted from the \mbox{\scshape{COLOR}} \normalfont simulations \citep{Hellwing_2016} in a $(70.4 \ \text{Mpc}/h)^3$ box with dark matter particle mass resolution $m_{\text{DM}} = 8.8 \times 10^6 \ M_{\odot}$. Therein, the SFR of a galaxy is assumed to be proportional to the gas accretion rate of its parent halo, $\dot{M}_{\text{gas}}$, normalised by a redshift-independent efficiency, $\epsilon(M_{\text{h}})$, of converting gas into stars,
\begin{equation}
\text{SFR}(M_{\text{h}}, z) = \epsilon(M_{\text{h}})\times \dot{M}_{\text{gas}}.
\end{equation}
The shortest timescale the model can probe, $\approx 0.05 \ t_{\text{H}}$, is $2-3$ times shorter than the dynamical timescale of the halo \citep[$\sim 0.15 \ t_{\text{H}}$,][]{Peebles_1980}. The model clearly misses many effects such as stellar feedback which can happen on even shorter timescales. At $z=7$, the shortest probed timescale is thus $0.05 \ t_{\text{H}} \sim 42$ Myr. We impose $M_{\star,\text{min}} = 2\times 10^7 \ M_{\odot}$ in accordance with \SBox.

\renewcommand{\arraystretch}{1.6}
\begin{table}
	\centering
	\caption{Galaxy formation models: (1) Name; (2) Cosmology, either Planck 2015 \citep{Planck_2015}, WMAP5 \citep{Komatsu_2009} or WMAP7 \citep{Komatsu_2011}; (3) Initial mass function (IMF), either Chabrier \citep{Chabrier_2003} or Salpeter \citep{Salpeter_1955}; (4) Model type.} 
	\noindent\begin{tabular}{L{1.6cm}L{1.6cm}L{1.6cm}L{1.7cm}}
    \mathrm{Name} & \mathrm{Cosmology} & \mathrm{IMF} & \text{Model Type}\\
	\midrule
	\hline
	\mathrm{TNG} & \text{Planck 2015} & \mathrm{Chabrier} & \text{Periodic Box} \\
	\mathrm{VELA} & \mathrm{WMAP5} & \mathrm{Chabrier} & \text{Zoom-In} \\
	\text{FirstLight} & \mathrm{WMAP5} & \mathrm{Chabrier} & \text{Zoom-In} \\
	\mathrm{EHM} & \mathrm{WMAP7} & \mathrm{Salpeter} & \text{Empirical Model} \\
	\bottomrule
	\end{tabular}
	\label{t_sims}
\end{table}
\renewcommand{\arraystretch}{1}

\subsection{FirstLight Simulations}
\label{ss_fl}
The \FL \ simulations \citep{Ceverino_2017} were run with the Adaptive Mesh Refinement (AMR) code ART \citep{Kravtsov_1997, Kravtsov_2003, Ceverino_2009}. Besides gravity and hydrodynamics, on sub-grid level the code incorporates gas cooling due to atomic hydrogen and helium, metal and molecular hydrogen cooling, photoionization heating by a constant cosmological UV background with partial self-shielding, star formation and feedback (thermal + kinetic + radiative), as described in \cite{Ceverino_2018}. The parent haloes were selected at $z\sim 5$ from $N$-body simulations with box sizes $10 \ \text{Mpc}/h$ and $20 \ \text{Mpc}/h$ such that their maximum circular velocity
$V_{\text{max}}$ lies between $50$ and $250$ km/s. This range excludes very massive and rare haloes with number densities lower than $\sim 3 \times 10^{-4} \ h^3$Mpc$^{-3}$, as well as small haloes in which galaxy formation is inefficient. The dark matter particle mass resolution is $m_{\text{DM}} = 10^4 \ M_{\odot}$. In this work, we perform a joint study of zoom-ins based on both box sizes $10 \ \text{Mpc}/h$ and $20 \ \text{Mpc}/h$, to which we will refer to as \FLBBox.

\subsection{VELA Simulations}
\label{ss_vela_sims}
This set of hydrodynamic simulations was likewise performed with the ART code, referred to as the \VELA \ runs. The main features can be summarised as follows. The Eulerian gas dynamics is followed using an adaptive mesh refinement (AMR) approach. The dark matter particle mass resolution is $m_{\text{DM}} = 8 \times 10^4 \ M_{\odot}$ while the AMR maximum resolution is $17 - 35$ pc at all times. In the circumgalactic medium (at the virial radius of the dark matter halo), the median resolution amounts to $\sim 500$ pc. The virial masses of the 29 galaxies are chosen to be in the range $M_{\text{vir}} = 2\times 10^{11} - 2\times 10^{12} M_{\odot}$ at $z=1$ about a median of $4.6\times 10^{11} M_{\odot}$. Beside gravity and hydrodynamics, the code includes many physical processes relevant for galaxy formation: gas cooling by atomic hydrogen and helium, metal and molecular hydrogen cooling, photoionization heating by the UV background with partial self-shielding, star formation, stellar mass loss, metal enrichment of the ISM and stellar feedback. Supernovae and stellar winds are implemented by local injection of thermal energy as described in \cite{Ceverino_2009, Ceverino_2010, Ceverino_2012}. Radiative stellar feedback is implemented at a moderate level \cite{Ceverino_2014}. AGN feedback and feedback associated with cosmic rays and magnetic fields are not implemented. Note that we do not impose a $M_{\star} = 10^7-10^9 \ M_{\odot}$ selection on \VELA \ and \FL \ galaxies when calculating SFR duty cycles and mini-quenching timescales in Secs. \ref{ss_duty_cycles} and \ref{ss_how_long}.\par 

\subsection{Model Differences}
\label{ss_model_diffs}
While \TNG \ aims to realistically capture the formation and evolution of the large-scale structure and the galaxies embedded therein, the simulations required significant computational resources \citep[\IBox \ alone took $18.62$ million CPU core hours,][]{Nelson_2018}. In contrast, EHM is much simpler and provides a flexible framework for modelling SFRs which can be calibrated to fit observational data. However, the evolution of individual galaxies cannot be followed with fine spatio-temporal resolution, a shortcoming that the zoom-in simulations \VELA \ and \FL \ address. In \mbox{Table \ref{t_sims}}, we compare the cosmologies and IMFs assumed in the four galaxy formation models. The value of $\Omega_{\text{m}}$ in the Planck 2015 cosmology ($\Omega_{\text{m}} = 0.3089$) is significantly higher than the WMAP5 ($\Omega_{\text{m}} = 0.258$) value, while WMAP7 ($\Omega_{\text{m}} = 0.273$) is in between the two. We estimate the global abundance of (mini-)quenched galaxies from \TNG \ and EHM, finding consistent results in Sec. \ref{ss_abundance} despite the differences in $\Omega_{\text{m}}$. Duty cycle estimates of individual galaxies in \VELA \ and \FL \ (both adopting WMAP5) depend more on subgrid modelling than cosmology and are found to be consistent with these global abundance estimates as well, see Sec. \ref{ss_duty_cycles}. While the Chabrier IMF provides a better fit to observations of low-mass stars and brown dwarfs in the Galactic disc than the Salpeter IMF \citep{Chabrier_2003}, neither is well-motivated at high redshift of $z>4$ where background radiation fields, gas temperatures and densities are considerably different \citep{Riaz_2021}.

\subsection{Averaging Timescales}
\label{ss_av_timescale}
Both simulations and observations consistently demonstrate that using shorter SFR averaging timescales results in a higher normalization and increased scatter of the main sequence, particularly at the low-mass end \citep{Schaerer_2013, Hayward_2014, Speagle_2014, Sparre_2015, Caplar_2019, Donnari_2019}. In observational studies, part of this effect can be attributed to a sampling bias towards stars in their young evolutionary stages or those located in regions with intense star formation activity. Our choice of $\Delta t = 10$ Myr for \TNG, \ \VELA \ and \FL is on the low end of timescales, yet is necessary to resolve SFHs of galaxies with high enough fidelity to infer mini-quenching timescales. For EHM, we choose the shortest timescale we can probe in this model, $\Delta t = 0.05 \ t_{\text{H}}$, as the natural SFR averaging timescale.

\begin{figure*}
\includegraphics[width=\textwidth]{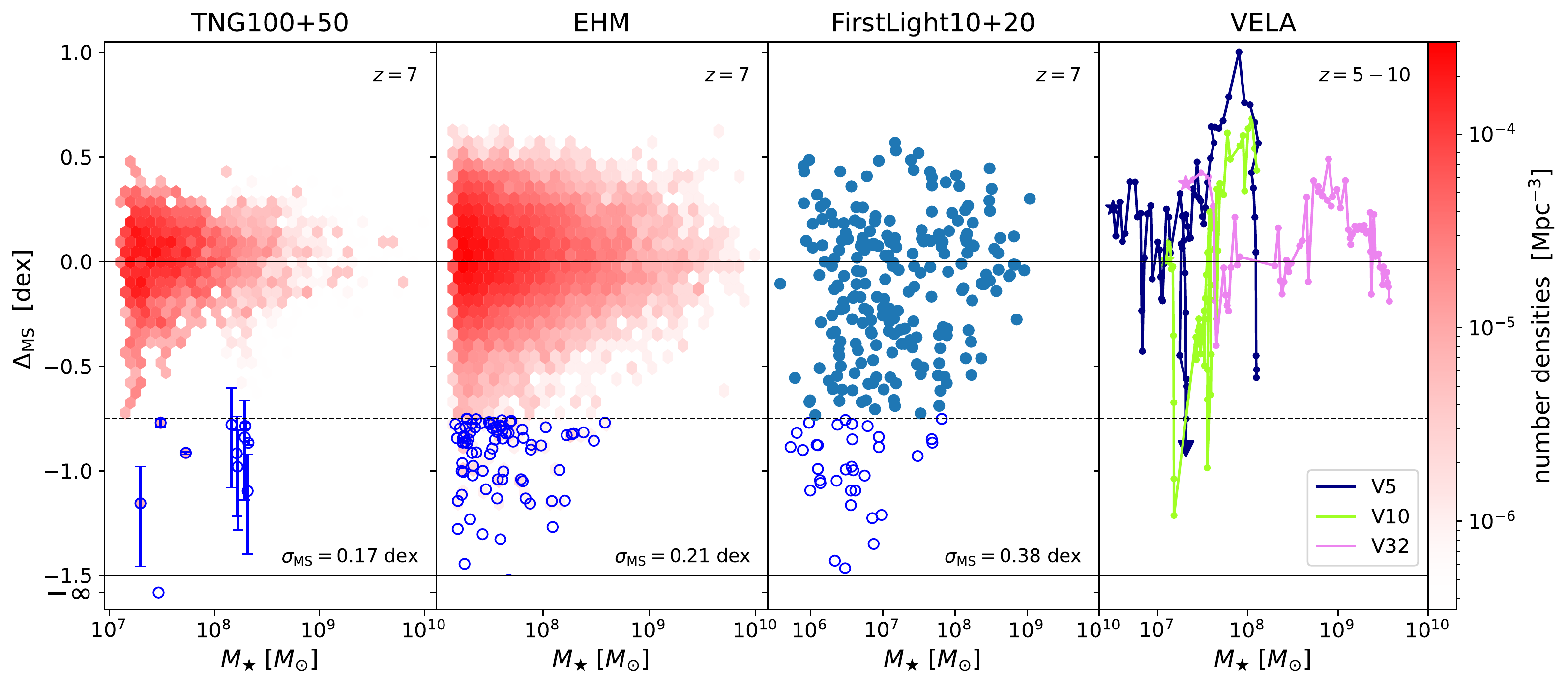}
\caption{Galaxies about the star-forming MS in a combined simulation study of \BBox \ (first panel, shown for $z=7$), EHM (second panel, shown for $z=7$), \FLBBox \ (third panel, shown for $z=7$) and \VELA \ (fourth panel, shown for $z=5-10$). Quiescent galaxies are highlighted with blue circles and lie below the selection threshold of $0.75$ dex below the MS, shown in dashed black. For \BBox, error estimates from bootstrapping over masses of newly formed stellar particles are added to each quiescent galaxy. Galaxies with $\text{SFR}=0$ are formally assigned a MS deviation of $\Delta_{\mathrm{MS}}=-\infty$. The colorbar indicates the comoving number density of galaxies. The total MS scatter averaged across $M_{\star} = 10^7-10^9 \ M_{\odot}$ is indicated in the bottom right of the panels. \VELA: Two galaxies with bursty SFHs (V5 and V10) and one galaxy with a more stable SFH (V32) are shown in different colours. SFRs are averaged over $10$ Myr, corresponding to the distance between two dots in the SFHs. Star symbols indicate the first star formation event ($z=10$ for V5 and V32 and $z=8$ for V10). The mini-quenching event of V5 (cf. Fig. \ref{f_seds}) is highlighted with a triangle symbol. The abundance of (mini-)quenched galaxies agrees across the models, with an overall fraction of $f_{\text{QG}} = 0.5-1.0$\% at $z=7$.}
\label{f_dms}
\end{figure*}

\subsection{Aperture Choice}
\label{ss_aperture}
Stellar mass and SFR of a galaxy depend on the radius within which they are computed. We aim to establish consistency with \citet[][]{Donnari_2019}, who have also demonstrated that too small apertures lead to an underestimate of the galaxy SFR, and \cite{Merlin_2019} as well as to mimic as accurately as possible the observational approach. To that end, in \TNG \ and \VELA, we use the values estimated within the 3D spherical galactocentric distance that corresponds to twice the stellar half mass radius $2\times R_{\mathrm{star,h}}$ of each galaxy. Recall that for \TNG, there is an additional gravitational boundedness criterion imposed under the hood by the SUBFIND algorithm. For \FL, galaxy stellar mass and SFRs are estimated within $0.15\ R_{\text{vir}}$ from the center of the parent halo, where $R_{\text{vir}}$ is the virial radius of the halo as per the spherical collapse result \citep{Bryan_1998}.

\subsection{Identification of the Star-Forming Main Sequence}
\label{ss_ms}
To define the star-forming MS in \TNG \ and EHM, we follow \cite{Donnari_2019} and iteratively remove quiescent galaxies until the median SFR in a given mass bin converges. The MS is thus allowed to deviate from a log-linear trend \citep[`bending MS',][]{Whitaker_2014, Donnari_2019}, as in fact it does by exhibiting a turnover at about $M_{\star} \gtrsim 10^{10.5} \ M_{\odot}$ \citep{Lee_2018, Tomczak_2016, Popesso_2023}. Depending on how a star-forming galaxy is defined observationally, this trend may persist up to $z\approx 6$. The bending MS estimation is performed at $z=4,5,6,7,8$ using stellar mass bins of width $0.25$ dex with results shown in Fig. \ref{f_ms}. We recover the well-known fact that the normalization increases with redshift at fixed stellar mass. This can be traced to the stellar mass growth time-scale (the ratio of stellar mass to star formation rate) which is expected to be comparable to the Hubble time \citep{Ma_2018, Ceverino_2018}. As seen in Fig. \ref{f_ms}, this trend becomes weaker towards higher redshift. To obtain SFHs in between these redshifts, we interpolate the MS.\par 

In \FLBBox \ and \VELA, the MS ridge is parametrized as
\begin{equation}
\text{sSFR}_{\text{MS}}(M_{\star},z)=s_{\text{b}}\left(\frac{M_{\star}}{10^{10}M_{\odot}}\right)^{\beta}(1+z)^{\mu}\ \text{Gyr}^{-1}.
\end{equation}
For \VELA, the best-fit parameters are $s_{\text{b}} = 0.046$, $\beta = 0.14$ and $\mu = 5/2$ \citep{Tacchella_2016} while for \FLBBox \ we find $s_{\text{b}} = 0.064$, $\beta = 0.05$ and $\mu = 2.45$. Across the mass range $M_{\star} = 10^7-10^{10} \ M_{\odot}$, the MS ridges are only consistent to within $\sim 30$\% between the four galaxy formation models at any given redshift, see Fig. \ref{f_ms}. Compared to observations of the star-forming MS as compiled by \cite{Popesso_2023}, while \BBox \ and \FLBBox \ are in good agreement therewith, EHM and \VELA \ typically have smaller normalizations. This deviation is evident for \VELA \ galaxies whose stellar-to-virial mass ratios are higher than deduced from observations \citep{Ceverino_2014}.
\par 

Quiescent (star-forming) galaxies are those whose SFR falls below (above) a certain relative distance from the median SFR at the corresponding mass. One popular choice is to define as quiescent those galaxies whose logarithmic specific star-formation rate is sSFR$< 10^{-11} \ \text{yr}^{-1}$ \citep{Donnari_2019, Merlin_2019}. Alternatively, some authors adopt a threshold such as $2\times \sigma_{\text{MS}} \sim 0.6$ dex \citep{Tacchella_2016} from the MS. However, here we choose a selection criterion of $0.75$ dex from the MS at all redshifts unless explicitly noted. This criterion should be seen as a necessary though not sufficient condition for mini-quenching. For \VELA \ and \FL \ galaxies whose SFHs we can resolve well, we further require that the quenching is only temporary. Mini-quenching timescales as well as duty cycles are quantified in Secs. \ref{ss_duty_cycles} and \ref{ss_how_long}, respectively.

\subsection{Calculating Spectral Energy Distributions}
\label{ss_seds}
To calculate a dust-free SED of a galaxy, we treat each stellar particle in the respective simulation as a simple stellar population (SSP) using a stellar population synthesis method. Here we opt for the Flexible Stellar Population Synthesis (FSPS) code \citep{Conroy_2009, ConroyGunn_2010} with MIST isochrones \citep{Paxton_2011, Paxton_2013, Paxton_2015, Choi_2016, Dotter_2016} and the MILES stellar library \citep{Sanchez_2006, Falcon_2011}, assuming a Chabrier initial mass function \citep{Chabrier_2003} consistent with the \TNG \ galaxy formation model.\par 

To correct for dust attenuation and extinction, we follow \cite{Nelson_2018}. When applied to \TNG, the model accurately reproduces the observed distribution of optical $(g-r)$ colors from the Sloan Digital Sky Survey. In addition to adopting a simple power-law extinction model \citep{Charlot_2000} for the attenuation by finite-lifetime birth clouds surrounding young stellar populations as well as the ambient diffuse ISM, we follow the distribution of metals and (neutral) hydrogen gas in and around each simulated galaxy. To obtain neutral hydrogen fraction estimates for star-forming cells in \TNG, we post-process the outputs and recalculate \cite[following][]{Villaescusa_2018_2} the equilibrium fractions according to the \cite{Springel_2003} model to account for the multiphase interstellar medium, including the presence of molecular hydrogen, $\text{H}_2$. For \VELA, we instead employ the total hydrogen column density to estimate the optical depth. The \cite{Nelson_2018} resolved dust model then attributes a neighborhood- and viewing angle-dependent attenuation to each stellar particle. We choose the viewing angle randomly as one of the vertices of the $N_s = 1$ \mbox{\scshape{HEALPIX}} \normalfont sphere \citep{Gorski_2005} oriented in simulation coordinates.

\renewcommand{\arraystretch}{1.6}
\begin{table*}
\centering
	\caption{MS scatter $\sigma_{\text{MS}}$ at various redshifts $z=4-8$, averaged across $M_{\star} = 10^7-10^9 \ M_{\odot}$ and decomposed into stellar mass bins of width $0.5$ dex. All estimates are given at redshifts $z=4,5,6,7,8$ for \BBox,  EHM and \FLBBox. Results at $z\leq 5$ for \FLBBox \ are not available. In accordance with quenched galaxy number densities in Fig. \ref{f_dms} (first two panels), we calculate a weighted standard deviation to avoid overcounting in the mass range $M_{\star}>2\times 10^8 \ M_{\odot}$ into which both \IBox \ and \SBox \ galaxies fall.}
\noindent\begin{tabular}{lL{0.7cm}L{0.7cm}L{0.7cm}L{0.7cm}L{0.7cm}L{0.7cm}L{0.7cm}L{0.7cm}L{0.7cm}L{0.7cm}L{0.7cm}L{0.7cm}L{0.7cm}L{0.7cm}L{0.7cm}}
\toprule
& \multicolumn{5}{c}{\BBox} & \multicolumn{5}{c}{EHM} & \multicolumn{5}{c}{\FLBBox}\\
\cmidrule(r){2-6}\cmidrule(l){7-11}\cmidrule(l){12-16}
$z$ & \text{Total} & 7-7.5  & 7.5-8  & 8-8.5 & 8.5-9 & \text{Total} & 7-7.5  & 7.5-8  & 8-8.5 & 8.5-9 & \text{Total} & 7-7.5  & 7.5-8  & 8-8.5 & 8.5-9\\
\midrule
\hline
$4$ & 0.26 & 0.28 & 0.24 & 0.23 & 0.22 & 0.32 & 0.33 & 0.31 & 0.31 & 0.31 & \text{n/a} & \text{n/a} & \text{n/a} & \text{n/a} & \text{n/a}\\
$5$ & 0.22 & 0.24 & 0.21 & 0.20 & 0.19 & 0.26 & 0.27 & 0.26 & 0.26 & 0.23 & 0.51 & 0.57 & 0.49 & 0.45 & 0.36\\
$6$ & 0.19 & 0.21 & 0.17 & 0.18 & 0.15 & 0.24 & 0.25 & 0.24 & 0.22 & 0.21 & 0.45 & 0.53 & 0.40 & 0.32 & 0.35\\
$7$ & 0.17 & 0.18 & 0.15 & 0.15 & 0.15 & 0.21 & 0.22 & 0.21 & 0.20 & 0.18 & 0.38 & 0.44 & 0.38 & 0.26 & 0.14\\
$8$ & 0.14 & 0.15 & 0.11 & 0.11 & 0.11 & 0.19 & 0.19 & 0.18 & 0.18 & 0.16 & 0.36 & 0.42 & 0.29 & 0.27 & 0.33\\
\bottomrule
\end{tabular}
\label{t_sigma_ms}
\end{table*}
\renewcommand{\arraystretch}{1}
\section{What is Mini-Quenching?}
\label{s_what_is_miniquenching}
By mini-quenching we refer to galaxies in the mass range $M_{\star} = 10^7-10^9 \ M_{\odot}$ \citep[such as the one reported by][]{Strait_2023,Looser_2023} that are temporarily quiescent. In this stellar mass range, star formation is bursty and regulated mainly by fluctuations in the gas inflow, stellar feedback and environmental effects such as gas-rich mergers. In \TNG, black holes are seeded when $M_{\text{DM}} = 5 \times 10^{10} \ h^{-1}M_{\odot}$ and AGN feedback is thus designed to be inefficient at the low-mass end \citep{Weinberger_2017}. However, overmassive black haloes in the centers of dwarfs can give rise to efficient AGN feedback at high redshift without violating observed HI gas mass constraints \citep{Koudmani_2022}, though direct observational evidence of such feedback channels is still needed. There is some contribution from photoheating in the presence of an ionizing background \citep{Okamoto_2008, Pawlik_2009, Brown_2014}, but \cite{Wu_2019} report that in self-consistent radiative transfer simulations (unlike our four galaxy formation models), photoheating due to reionization can suppress SFRs by more than $50$\% only in low-mass haloes, specifically $M_h < 10^{8.4} \ M_{\odot}$ at $z=6$. In this work, we do not investigate such reionization quenching at very low masses.\par 

\cite{Ma_2018} find that $M_{\star} \sim 10^8 \ M_{\odot}$ is the (weakly redshift dependent) mass transition threshold where SFHs begin to transition from bursty to stable. Note that the physical mechanisms underlying the transition from bursty to steady star formation are complex \citep{Sparre_2017, Hopkins_2023, Gurvich_2023} and might be related to the virialization of the inner circumgalactic medium \citep{Stern_2021}.\par

As mentioned in the introduction, mini-quenching does not refer to long-term quenching of galaxies with $M_{\star}>10^{10} \ M_{\odot}$. This form of quenching of high-mass galaxies is commonly observed at lower redshift, but see \cite{Glazebrook_2017, Forrest_2020, Valentino_2020, Nanayakkara_2022, Carnall_2023}. The primary factor governing star formation within high-mass systems is likely radio mode feedback from supermassive black holes, observable via radio lobes, X-ray cavities and radio-optical correlations \citep{Kormendy_2013, Terrazas_2020, Houston_2023}. \par

In the following, we compare various statistics between \TNG, EHM, \FL \ and \VELA \ to demonstrate that their predictions for high-redshift mini-quenching are in good agreement with each other.

\subsection{Scatter around the MS}
\label{ss_scatter_ms}
The distribution of galaxies around the star-forming MS (see Sec. \ref{ss_ms}) in the various galaxy formation models is shown in Fig. \ref{f_dms}. Quiescent galaxies suffer from a higher SFR uncertainty, which we accommodate by adding error estimates from bootstrapping to the \BBox \ results. Specifically, for each galaxy we bootstrap over the initial masses of stellar particles that formed in the last $10$ Myr (averaging timescale) before $z=7$. We assume a scatter of $m_{\star}/2$ if only $1$ stellar particle of mass $m_{\star}$ formed over the last $10$ Myr. Note that upper and lower error bars are asymmetric.\par 

We find that the scatter $\sigma_{\text{MS}}$ around the MS (as quantified in Table \ref{t_sigma_ms}) decreases towards higher stellar mass\footnote{We note that the scatter around the MS is moderately inconsistent across the \TNG \ simulations. When restricting to the $1.5-3\times \ 10^8 M_{\odot}$ range at $z=7$, \mbox{\scshape{TNG100-1}} \normalfont exhibits a RMS scatter of $0.21$ dex in contrast to $0.12$ dex in \mbox{\scshape{TNG50-1}} \normalfont and $0.15$ dex in \mbox{\scshape{TNG50-2}}. \normalfont Since \mbox{\scshape{TNG100-1}} \normalfont resolves galaxies poorly in this mass range while the \SBox \ scatter values are consistent with each other, some \mbox{\scshape{TNG100-1}} \normalfont galaxies are at large negative $\Delta_{\text{MS}}$ because of poor resolution.}. For instance, at $z=7$ \BBox \ galaxies in the stellar mass range $M_{\star} = 10^7-10^{7.5} \ M_{\odot}$ have a scatter of $\sigma_{\text{MS}} = 0.18$ while those of mass $M_{\star} = 10^{8.5}-10^9 \ M_{\odot}$ have $\sigma_{\text{MS}} = 0.15$. This trend is in agreement with results from \mbox{\scshape{FIRE-2}} \normalfont \citep{Ma_2018} and is a result of bursty star formation regulated mainly by stellar feedback and environmental effects. However, the decrease of the scatter towards higher mass is less pronounced in \mbox{\scshape{Flares}} \normalfont simulations \citep{Lovell_2022}, which resolve galaxies above $M_{\star} \sim 10^9 \ M_{\odot}$. \mbox{\scshape{Flares}} \normalfont employs a physically motivated model for AGN feedback that takes into account the dynamics of the accretion disk and the surrounding gas as well as the radiation emitted by the AGN, which better reproduces the observed properties of massive galaxies at high redshifts such as sizes, masses and stellar populations \citep{Vijayan_2020, Roper_2022}.\par 
\renewcommand{\arraystretch}{1.6}
\begin{table*}
\centering
	\caption{Fraction and abundance of quenched galaxies (QGs) in the stellar mass range $M_{\star} = 10^7-10^9 \ M_{\odot}$: (1) fraction of quenched galaxies in percent, with error estimates from bootstrap resampling; (2) fraction of environmentally mini-quenched galaxies (EMQGs) among all quenched galaxies in percent; (3) comoving number density of quenched galaxies in Mpc$^{-3}$. All estimates are given at redshifts $z=4,5,6,7,8$ for \IBox, \SBox, a joint analysis of \IBox \ and \SBox \ as well as EHM. The fraction of EMQGs cannot be calculated within EHM and is thus not shown.}
\noindent\begin{tabular}{lL{1.1cm}L{1.1cm}L{1.1cm}L{1.1cm}L{1.1cm}L{1.1cm}L{1.1cm}L{1.2cm}L{1.2cm}L{1.2cm}L{1.2cm}}
\toprule
& \multicolumn{4}{c}{fraction of QGs $[\%]$} & \multicolumn{3}{c}{\# EMQGs / \# QGs $\times \ 100$ $[\%]$} & \multicolumn{4}{c}{\# density of QGs [Mpc$^{-3}$]}\\
\cmidrule(r){2-5}\cmidrule(l){6-8}\cmidrule(l){9-12}
$z$ & \text{T100} & \text{T50}  & \text{T100+50}  & \text{EHM} & \text{T100} & \text{T50}  & \text{T100+50} & \text{T100} & \text{T50}  & \text{T100+50} & \text{EHM}\\
\midrule
\hline
$4$ & 4.2\pm 0.2 & 3.6\pm 0.3 & 4.0\pm 0.4 & 2.5\pm 0.2 & 25.5 & 14.1 & 18.8 & 3.2\times 10^{-4} & 2.6\times 10^{-3} & 5.3\times 10^{-4} & 2.2\times 10^{-3}\\
$5$ & 2.2\pm 0.3 & 1.6\pm 0.4 & 1.9\pm 0.3 & 1.2\pm 0.1 & 18.2 & 26.4 & 21.8 & 8.5\times 10^{-5} & 6.9\times 10^{-4} & 1.4\times 10^{-4} & 7.3\times 10^{-4}\\
$6$ & 1.0\pm 0.5 & 0.8\pm 0.3 & 0.9\pm 0.2 & 0.9\pm 0.1 & 6.3 & 30.0 & 15.4 & 1.5\times 10^{-5} & 2.0\times 10^{-4} & 3.2\times 10^{-5} & 3.1\times 10^{-4}\\
$7$ & 1.0\pm 0.8 & 0.3\pm 0.4 & 0.6\pm 0.3 & 0.5\pm 0.2 & 12.5 & 66.7 & 27.3 & 5.9\times 10^{-6} & 3.6\times 10^{-5} & 8.7\times 10^{-6} & 9.1\times 10^{-5}\\
$8$ & 0.0\pm 0.0 & 0.0\pm 0.0 & 0.0\pm 0.0 & 0.3\pm 0.2 & 0.0 & 0.0 & 0.0 & 0.0 & 0.0 & 0.0 & 2.5\times 10^{-5}\\
\bottomrule
\end{tabular}
\label{t_mqg}
\end{table*}
\renewcommand{\arraystretch}{1}

The most common approach to infer the evolution of the star-forming MS observationally is by combining galaxy samples at different redshifts for which the SFRs are estimated for every galaxy. More robust determinations of total SFRs and the portions associated with the unobscured, SFR$_{\text{FUV}}$, and obscured, SFR$_{\text{IR}}$, regimes bridge the gap with other constraints such the galaxy stellar mass function of star-forming galaxies and the FUV and IR luminosity functions \citep{Puebla_2020}. UV-derived SFRs tend to indicate a scatter of $\sim 0.25-0.30$ dex \citep{Elbaz_2007, Whitaker_2012, Speagle_2014, Fang_2018}. Driven by stochastic, bursty SFHs at the low-mass end and the presence of bulges, bars but also AGN at the high-mass end, the dispersion can increase to $\sim 0.4$ dex \citep{Santini_2017, Popesso_2019, Guo_2015}. According to \cite{Guo_2015, Willett_2015, Davies_2019}, the scatter might follow a minimum vertex parabolic `U'-shape decreasing with stellar mass from $\log (M_{\star}/M_{\odot}) \sim 8-10$ and then increasing at $\log (M_{\star}/M_{\odot}) \sim 10-11.5$. Note though that few studies go above $z\approx 5$.\par

\TNG \ likely underestimates the scatter around the MS at higher mass compared to observations given that the AGN feedback is modeled as a simplified subgrid process, where the (fully isotropic) energy and momentum injection from the AGN is regulated by a set of rules based on the accretion rate and black hole mass \citep{Weinberger_2017}. The EHM, \VELA, \ \mbox{\scshape{FIRE-2}} \normalfont and \FL \ models do not include feedback from AGN and likewise cannot fully capture the complex feedback processes in massive galaxies.\par 

For EHM (Fig. \ref{f_dms}, second panel), we in addition see that the overall number of galaxies is larger than in the case of \BBox \ ($17226$ vs $2369$ at $z=7$ in the range $M_{\star} = 10^7-10^9 \ M_{\odot}$). EHM might indeed overpredict the galaxy stellar mass function at the low-mass end as hinted at by \cite{Tacchella_2018} whilst the NIR band luminosity function and by extension the galaxy stellar mass function predicted by \TNG \ are largely consistent with observations at high redshift \citep{Shen_2022}. Despite that, the total EHM scatter $\sigma_{\text{MS}} = 0.21$ dex around the MS agrees well with the scatter $\sigma_{\text{MS}} = 0.17$ dex found for \BBox, see Table \ref{t_sigma_ms}. The MS scatter in EHM comes directly from the scatter in the halo mass accretion rate. \FLBBox \ exhibits a higher MS scatter $\sigma_{\text{MS}} = 0.38$ dex than \TNG \ and EHM, which can be traced to \FL \ being able to resolve intense SF bursts on timescales smaller than $\sim 10$ Myr \citep[also see][]{Ceverino_2018}.\par 
In Table \ref{t_sigma_ms}, we also quantify the redshift evolution of the total and stellar mass bin-decomposed MS scatter. \BBox, \ EHM and \FLBBox \ galaxies all exhibit a larger scatter towards lower redshift. This is in agreement with \cite{Ma_2018, Caplar_2019} and is a result of hierarchical structure formation which leads to a more diverse range of galaxy properties at later epochs. At lower redshift, we also probe a wider range of large-scale environments such as galaxy clusters, which contributes to the increased scatter in the star-forming MS.

\subsection{Abundance of (Mini-)Quenching Events}
\label{ss_abundance}
How common are (mini-)quenching events at high redshift $z=4-8$? While each of the four galaxy formation models probes a slightly different stellar mass range, we find that estimates from all four are largely consistent with a picture in which the population first appears below $z\approx 8$, after which the fraction of (mini-)quenched galaxies increases with cosmic time, from $\sim 0.5-1.0$\% at $z=7$ to $\sim 2-4$\% at $z=4$, corresponding to comoving number densities of $10^{-5}$ Mpc$^{-3}$ and $10^{-3}$ Mpc$^{-3}$, respectively.\par
\begin{figure}
\includegraphics[width=\linewidth]{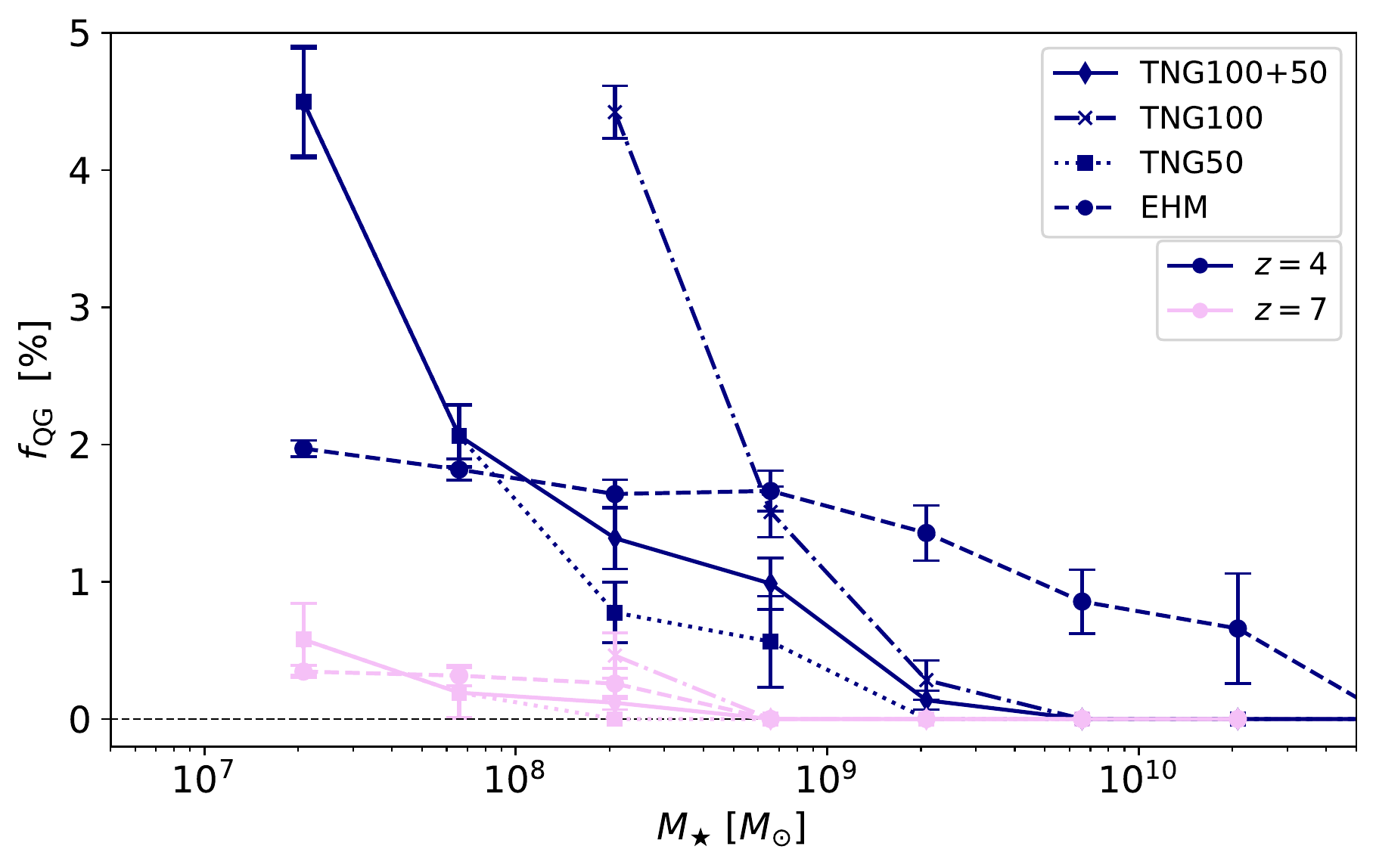}
\caption{Quenched galaxy fraction $f_{\text{QG}}$ vs stellar mass $M_{\star}$ at various redshifts (see legend) in a combined simulation study of \BBox \ (solid), in the individual boxes \IBox \ (dash-dotted), \SBox \ (dotted) and EHM (dashed). The population of quenched galaxies first appears below $z\approx 8$, after which their fraction increases with cosmic time, from $\sim 0.5-1.0$\% in the mini-quenching mass range $M_{\star}=10^7-10^9 \ M_{\odot}$ at $z=7$ to $\sim 2-4$\% at $z=4$. The error bar on each estimate denotes the standard deviation as obtained with bootstrap resampling. The overall mass-agnostic quenched galaxy fraction is shown in \mbox{Table \ref{t_mqg}}. There is qualitative agreement between \TNG \ and the simplistic EHM approach, even though the latter does not model stellar feedback.}
\label{f_abundance}
\end{figure}
Fig. \ref{f_abundance} compares the fraction of quenched galaxies in \TNG \ and EHM. A combined analysis of \BBox \ yields an overall quenched galaxy fraction of $f_{\text{QG}} = 0.6$\% at $z=7$, corresponding to a comoving number density of $8.7\times 10^{-6}$ Mpc$^{-3}$. For a comparison of redshifts $z=4,5,6,7,8$ see Table \ref{t_mqg}. For \IBox, we find that at $z=7$ an overall $f_{\text{QG}} = 1.0$\% of galaxies are mini-quenched ($8$ out of $826$ galaxies), translating\footnote{Note that this value is higher by several dex than the value obtained with a simple extrapolation from a low-redshift \IBox \ analysis reported by \cite{Merlin_2019}. The discrepancy can be traced back to our lower stellar mass limit of $M_{\star} > 2\times 10^8 \ M_{\odot}$ rather than $M_{\star} > 5\times 10^9 \ M_{\odot}$ and the selection threshold ($0.75$ dex rather than sSFR$< 10^{-11} \ \text{yr}^{-1}$).} to a comoving number density of $5.9\times 10^{-6}$ Mpc$^{-3}$. In \SBox, the fraction of lowest-mass quenched galaxies around $M_{\star} \sim 2\times 10^7 \ M_{\odot}$ at $z=7$ reads $0.7$\% while the overall one is $f_{\text{QG}} = 0.3$\% ($4$ out of $1543$ galaxies). The comoving number density is thus $3.6\times 10^{-5}$ Mpc$^{-3}$, which is higher than the \IBox \ value since the minimum stellar mass $M_{\star,\text{min}} = 2\times 10^7 \ M_{\odot}$ is lower.

\renewcommand{\arraystretch}{1.6}
\begin{table}
	\centering
	\caption{SFR duty cycles in \VELA \ and \FLBBox: (1) epoch of observation $z$; (2) selection threshold in dex below the MS; (3) median and $16 / 84$ percentile of the SFR duty cycle $f_{\text{duty}}$ in percent.} 
	\noindent\begin{tabular}{lL{1.4cm}L{1.4cm}L{1.4cm}L{1.4cm}}
	\toprule
	& \multicolumn{2}{c}{\VELA} & \multicolumn{2}{c}{\FLBBox}\\
	\cmidrule(r){2-3}\cmidrule(r){4-5}
	$z$ & -0.75 \ \text{dex} & -0.6 \ \text{dex}& -0.75 \ \text{dex} & -0.6 \ \text{dex}\\
	\midrule
	\hline
	$4$ & 98.4^{+1.6}_{-6.8}\ \% & 96.1^{+2.3}_{-8.4}\ \% & 94.7^{+5.3}_{-9.6}\ \% & 91.7^{+8.3}_{-10.0}\ \%\\
	$5$ & 98.4^{+1.6}_{-2.6}\ \% & 96.3^{+3.5}_{-2.5}\ \% & 91.0^{+9.0}_{-16.2}\ \% & 86.0^{+14.0}_{-17.9}\ \%\\
	$6$ & 98.2^{+1.8}_{-3.1}\ \% & 95.9^{+4.1}_{-3.6}\ \% & 90.8^{+9.2}_{-25.0}\ \% & 83.6^{+16.4}_{-26.3}\ \%\\
	$7$ & 98.6^{+1.4}_{-4.1}\ \% & 95.8^{+4.2}_{-4.8}\ \% & 99.9^{+0.1}_{-37.9}\ \% & 92.4^{+7.6}_{-42.6}\ \%\\
	$8$ & 99.1^{+0.9}_{-5.6}\ \% & 97.4^{+2.6}_{-9.5}\ \% & 99.9^{+0.1}_{-0.0}\ \% & 99.9^{+0.1}_{-0.0}\ \%\\
	\bottomrule
	\end{tabular}
	\label{t_duty_cycle}
\end{table}
\renewcommand{\arraystretch}{1}
\renewcommand{\arraystretch}{1.6}
\begin{table}
	\centering
	\caption{Mini-quenching timescales $\tau_{\mathrm{MQ}}$ in \VELA \ and \FLBBox \ averaged over $z=4-8$: (1) selection threshold in dex below the MS; (2) median and $16 / 84$ percentile of $\tau_{\mathrm{MQ}}$ in Myr.} 
	\noindent\begin{tabular}{lL{2.1cm}L{2.1cm}}
	\toprule
	dex & \VELA & \FLBBox \\
	\midrule
	\hline
	-0.75 & 16.5^{+23.3}_{-7.7}\ \text{Myr} & 35.1^{+41.8}_{-24.5}\ \text{Myr}\\
	-0.6 & 21.5^{+34.4}_{-18.2}\ \text{Myr} & 42.4^{+42.7}_{-32.0}\ \text{Myr}\\
	\bottomrule
	\end{tabular}
	\label{t_tau_mq}
\end{table}
\renewcommand{\arraystretch}{1}

Since the dependence on stellar mass is less pronounced in EHM and \IBox \ than in \SBox, Fig. \ref{f_abundance} also demonstrates that EHM (based on \mbox{\scshape{COLOR}} \normalfont merger trees) and \IBox \ explore a different dynamic range than \SBox, and better sample the intermediate-mass range of the galaxy stellar mass function. Since EHM might overpredict the galaxy stellar mass function at the low-mass end (cf. Sec. \ref{ss_scatter_ms}), quenched galaxy number densities are higher by about $1/2$ dex than in the combined \BBox \ analysis, see Table \ref{t_mqg}.

\subsection{Duty Cycles in \FL \ \textbf{and \VELA}}
\label{ss_duty_cycles}
It is straightforward to look at the evolution of selected zoom-in galaxies around the ridge of the star-forming MS. In Fig. \ref{f_dms} (right panel), we show SFHs of two \VELA \ galaxies with bursty SFHs (V5 and V10) and one galaxy with a more stable SFH (V32). However, extracting MQG number densities from zoom-ins is ill-defined\footnote{If zoom-in simulations are specially designed such that galaxies are sampled from a range of overdensities in the parent simulation, it is possible to weight the regions to produce composite distribution functions, including MQG number densities, see \cite{Lovell_2022}.}. How do we know whether the abundance of MQGs in \FL \ and \VELA \ is consistent with \TNG \ and EHM?\par

To this end, it is illustrative to investigate the SFR duty cycle. It denotes the fraction of time spent in an active phase, i.e., the ratio between the actively star forming ($\Delta_{\text{MS}} > [-0.75, -0.6]$ dex) time interval $\Delta t_{\text{on}}$, and the time elapsed between the first star formation event $t_{\text{form}}$ and the epoch of observation $t_{\text{obs}}$ \citep[see][]{Gelli_2023},
\begin{equation}
f_{\text{duty}} = \frac{\Delta t_{\text{on}}}{t_{\text{obs}}-t_{\text{form}}}.
\end{equation}

Table \ref{t_duty_cycle} presents a comparison of duty cycle estimates for redshifts $z=4-8$, with two different selection thresholds $\Delta_{\text{MS}} = [-0.75, -0.6]$ dex. Notably, \VELA \ galaxies demonstrate an active star formation phase for $\sim 96.1 - 98.4$\% of their time at $z=4$. The duty cycle increases with the epoch of observation $z$ and at $z=8$ attains values $\sim 99$\%.\par 

In contrast, \FLBBox \ galaxies display systematically lower values of $f_{\text{duty}}$ compared to \VELA. At $z=5$, the duty cycle drops as low as $\sim 91$\%. Some of this discrepancy can be attributed to \FL \ galaxies benefiting from an eight-fold increase in resolution compared to \VELA \ galaxies. This enhanced resolution allows \FL \ to resolve many lower-mass systems, which tend to exhibit more bursty star formation behavior (see Fig. \ref{f_dms}). However, even when focusing the $f_{\text{duty}}$ analysis on galaxies within a specific mass range, the lower tail for \FL \ galaxies remains substantial. For instance, at $z=6$, when restricting the analysis to galaxies with $M_{\star} \in [10^8-10^9] \ M_{\odot}$, the duty cycle is estimated to be $98.5^{+0.5}_{-0.8}$\% for \VELA \ and $99.9^{+0.1}_{-15.5}$\% for \FLBBox. This indicates that the difference in duty cycle between \FLBBox \ and \VELA \ remains significant, even within the confines of a specific mass range where both \VELA \ and \FLBBox \ galaxies are well resolved. While the duty cycle of a galaxy also depends on large-scale environments (see Sec. \ref{ss_drivers}), we thus conclude that the stronger feedback in \FL \ \citep[see][]{Ceverino_2018} compared to \VELA \ is also reflected in lower duty cycles.\par 

\begin{figure}
\includegraphics[width=0.5\textwidth]{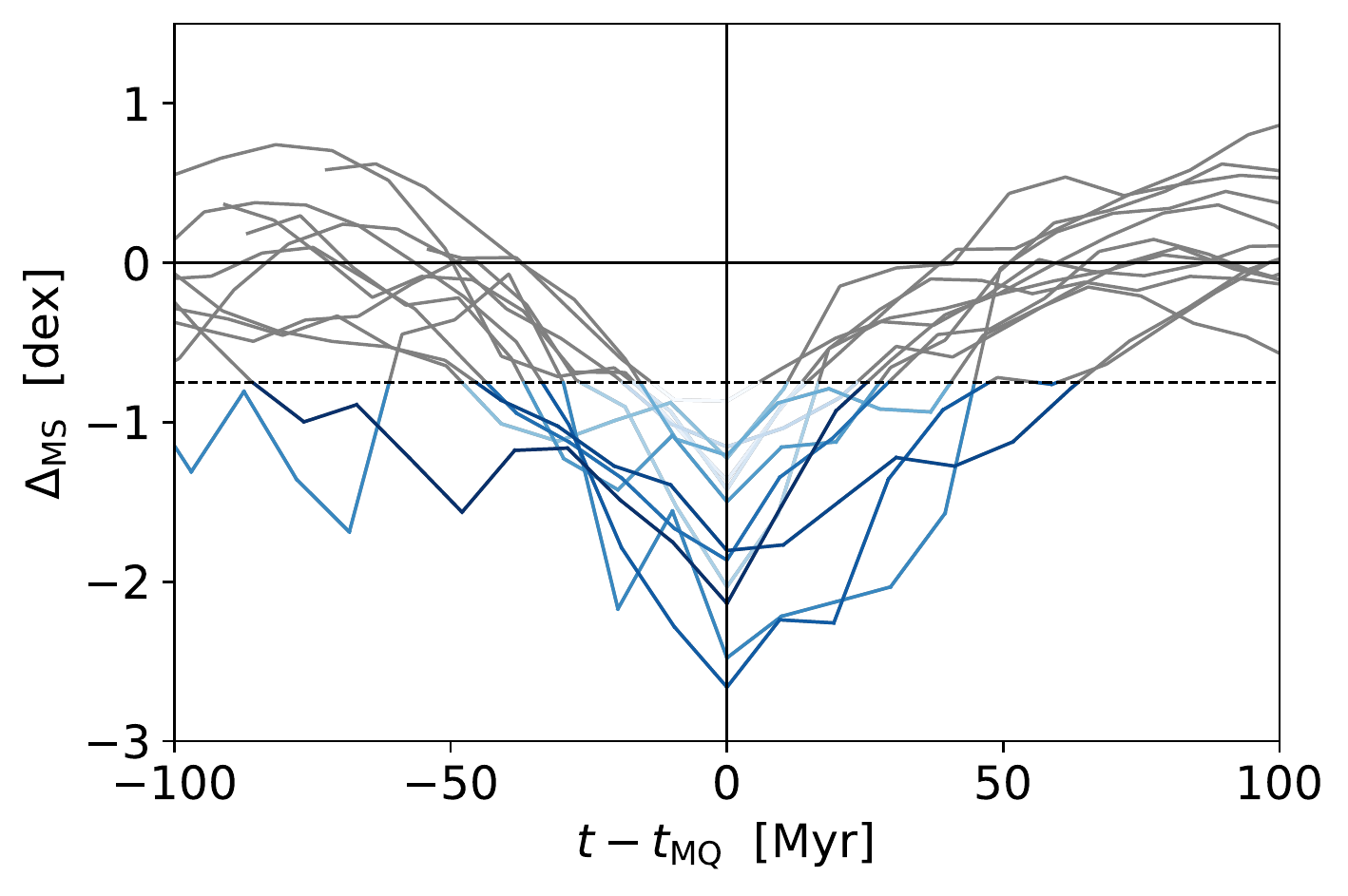}
\caption{Bursty SFHs of \FLBBox \ galaxies around $12$ selected mini-quenching events in the redshift range $z=4-8$. Each trajectory shows the deviation from the MS from $100$ Myr before until $100$ Myr after the mini-quenching event. One galaxy undergoes two mini-quenching events separated by $\sim 30$ Myr. The selection threshold of $0.75$ dex below the MS is shown in dashed black. Mini-quenching timescales $\tau_{\text{MQ}}$ in Table \ref{t_tau_mq} are obtained by summing up the time spent below this threshold. We color-code trajectories by the value of $\tau_{\text{MQ}}$: the higher $\tau_{\text{MQ}}$, the darker the blue shade.}
\label{f_t_minus_tmq}
\end{figure}
The SFR duty cycle estimates are in good agreement with MQG number densities obtained for \TNG \ (see Table \ref{t_mqg}). At $z=7$, $0.6$\% of galaxies are mini-quenched in a joint analysis of \IBox \ and \SBox. Assuming this fraction remains constant at all $z>7$, this would correspond to a duty cycle of $f_{\text{duty}} = 99.4\%$ compared to the value $f_{\text{duty}} = 98.6^{+1.4}_{-4.1}\%$ we infer for \VELA \ and $f_{\text{duty}} = 99.9^{+0.1}_{-37.9}\%$ for \FLBBox. However, the $f_{\text{duty}}$ values inferred are not consistent with those from the \mbox{\scshape{SERRA}} \normalfont simulations \citep[$f_{\text{duty}} \sim 0.60-0.99$,][]{Pallottini_2022, Gelli_2023}, a discrepancy which we will elaborate on in Sec. \ref{ss_discussion}.

\subsection{How Long is a Mini-Quenching Event?}
\label{ss_how_long}
In the compaction-triggered quenching model \citep{Dekel_2014, Zolotov_2015, Lapiner_2023}, galaxies at $z\approx 2-3$ undergo three evolutionary phases: cold gas accretion, compaction and post-compaction, and quenching. Before this final successful quenching attempt, galaxies oscillate about the MS ridgeline on timescales of $\sim 0.4 \ t_{\text{H}}$, as found by \cite{Tacchella_2016} in the \VELA \ simulation suite at $z=2-4$. \cite{Houston_2023} found that the oscillation period increases towards higher stellar mass galaxies to about $1.98 \pm 2.27$ Gyr for galaxies with $M_{\star} = 10^{11.5}-10^{12} \ M_{\odot}$.\par

Here we are interested in a different but related timescale. Given a SFH, we estimate the accumulated time each MQG spends $0.75$ dex or further below the MS, which we define as the \textit{mini-quenching timescale} $\tau_{\mathrm{MQ}}$.\par 

Many SFHs of MQGs in \TNG \ at high redshift trace out an irregular pattern (especially) around the mini-quenching event ($\pm 20$ Myr), suggesting a bursty mode of star formation. However, SFHs are poorly resolved, with many displaying a distinctive V-shaped feature around the mini-quenching event. We thus only capture a glimpse into said event and cannot put meaningful constraints on $\tau_{\mathrm{MQ}}$. A similar challenge is encountered with EHM which is based on halo merger trees extracted from the \mbox{\scshape{COLOR}} \normalfont simulations in a $(70.4 \ \text{Mpc}/h)^3$ box. By tracking the main progenitor and main descendant branch of halos, \cite{Tacchella_2018} showed that in this empirical framework many galaxies oscillate around the MS at $z>4$, hinting at bursty star formation as a result of halo mergers. However, due to limitations in resolution, EHM does not provide a robust framework for estimating $\tau_{\mathrm{MQ}}$ either.\par

Instead, we investigate mini-quenching timescales from our zoom-ins \FL \ and \VELA, see Table \ref{t_tau_mq}. For a total of $20$ mini-quenching events in the redshift range $z=4-8$, we find a median and $16 / 84$ percentile of $\tau_{\mathrm{MQ}} = 16.5^{+23.3}_{-7.7}$ Myr for \VELA. For comparison, when imposing a selection threshold of $0.6 \ \text{dex} \sim 2\times \sigma_{\text{MS}}$ \citep{Speagle_2014}, we find $\tau_{\mathrm{MQ}} = 21.5^{+34.4}_{-18.2}$ Myr. In these distributions, \VELA \ galaxy \scshape{V28} \normalfont constitutes an outlier which stays quiescent for $\sim 220$ Myr in the redshift window $z\approx 4.9-5.7$.\par 

For \FLBBox \ galaxies, we show SFHs around $12$ selected mini-quenching events in Fig. \ref{f_t_minus_tmq}. The well-resolved trajectories demonstrate the bursty, stochastic nature of star formation at high redshift. Mini-quenching events typically last longer than in \VELA \ as expected from the lower SFR duty cycles. The distribution is centered around $\tau_{\mathrm{MQ}} \sim 40$ Myr with a heavy upper tail. In fact, the longest mini-quenching events can last up to $\tau_{\mathrm{MQ}} \sim 100$ Myr before the galaxy rejuvenates.\par 

Expressed in terms of the Hubble time $t_{\text{H}}(z=7) \sim 760$ Myr, the typical timescale at $z=7$ is $\tau_{\mathrm{MQ}} \sim 0.02 \ t_{\text{H}}$. This timescale is more than an order of magnitude smaller than the oscillation timescale $\sim 0.4 \ t_{\text{H}}$ about the ridge of the MS, indicating that the SFR needs to change significantly on shorter timescales around the mini-quenching event (with a more steady evolution closer to the MS). In the framework of correlated stochastic processes \citep{Kelson_2014,Caplar_2019,Abramson_2020,Tacchella_2020,Iyer_2020,Iyer_2022}, this might suggest different amounts of power on different temporal scales.\par

The mini-quenching timescale $\tau_{\mathrm{MQ}} \sim 20-40$ Myr is several times ($\sim 1/6$) shorter than the overall free-fall time $t_{\mathrm{ff}} = (3\pi / (32G\rho))^{1/2} \sim 240$ Myr of a halo of mass $M_{\text{h}} = 6 \times 10^9\ M_{\odot}$, hosting a galaxy such as \scshape{V5} \normalfont of mass $M_{\star} = 2.1 \times 10^7 \ M_{\odot}$ (see Sec. \ref{s_sed_jades}) based on the observed stellar-to-halo-mass relation \citep{Girelli_2020}. However, we argue that $\tau_{\mathrm{MQ}}$ is close to the local free-fall timescale $t_{\mathrm{ff,loc}}$ of the \textit{inner halo} where galaxies typically reside. In this picture, after the expulsion of gas in the wake of stellar feedback the galaxy finds itself in a state of mini-quenching until gas gets reaccreted/falls onto the galaxy. The mini-quenching timescale is influenced by the mass distribution of the halo, the density and temperature of the gas, and feedback effects such as radiative winds \citep{Stern_2021, Gelli_2023}. At high densities and low metallicities, $t_{\mathrm{ff,loc}}$ can be shorter than the stellar feedback timescales \citep{Dekel_2023}, in which case $\tau_{\mathrm{MQ}}$ will be the closest to $t_{\mathrm{ff,loc}}$.

\subsection{What Drives Mini-Quenching?}
\label{ss_drivers}
The fluctuation of SFRs from a state of mini-quenching to an episodic burst is typically mediated by the interplay between gas-rich mergers and the steady influx of cold gas streams, periodically inhibited by the feedback from evolved stars. EHM captures mergers insofar that an increased dark matter accretion rate is assumed to result in enhanced gas accretion. \TNG, \FL \ and \VELA \ take account of stellar feedback and thus capture bursty star formation processes on smaller temporal scales.\par 

Even though the effects of ram pressure and virial shocks are less efficient at high redshift \citep{Fujita_2001, Birnboim_2003, Maier_2021}, in \TNG \ we find that tidal interactions during close galaxy-galaxy encounters (which are more common at high redshift than today) play an important role in determining SFRs of (especially) low-mass galaxies. We follow the main progenitor and main descendant branches of each galaxy\footnote{The main progenitor and main descendant branches are obtained from merger trees constructed at the subhalo level using the \mbox{\scshape{SubLink}} \normalfont algorithm \citep{Rodriguez_Gomez_2015}.} around the mini-quenching event, and search for discontinuities in the stellar mass evolution which are not matched by a corresponding change in SFRs. If said discontinuity occurs right before or after the mini-quenching event, we label the galaxy an \textit{environmentally mini-quenched} galaxy (EMQG). We refrain from adopting a more rigorous merger-only selection based on merger trees since we find that several quenched galaxies are tidally distorted (sometimes with an accompanying \textit{decrease} of stellar material falling inside the aperture) without fully merging.\par 

The fraction of EMQGs among all quenched galaxies (cf. Table \ref{t_mqg}) is highest for galaxies on the low-mass end (\SBox) and increases with redshift (up to $\# \ \text{EMQG} / \# \ \text{QG} \times 100 \sim 66.7\%$ at $z=7$). For intermediate-mass galaxies sampled from \IBox, the redshift trend is reversed, and peaks at $z=4$ with $25.5\%$. In a combined analysis of \BBox, we find an EMQG fraction of $27.3\%$. Tidally induced mini-quenching events such as complete and incomplete mergers thus play an important role in the context of bursty star formation.

\begin{figure*}
\begin{subfigure}{0.48\textwidth}
\hspace{-0.7cm}
\includegraphics[scale=0.6]{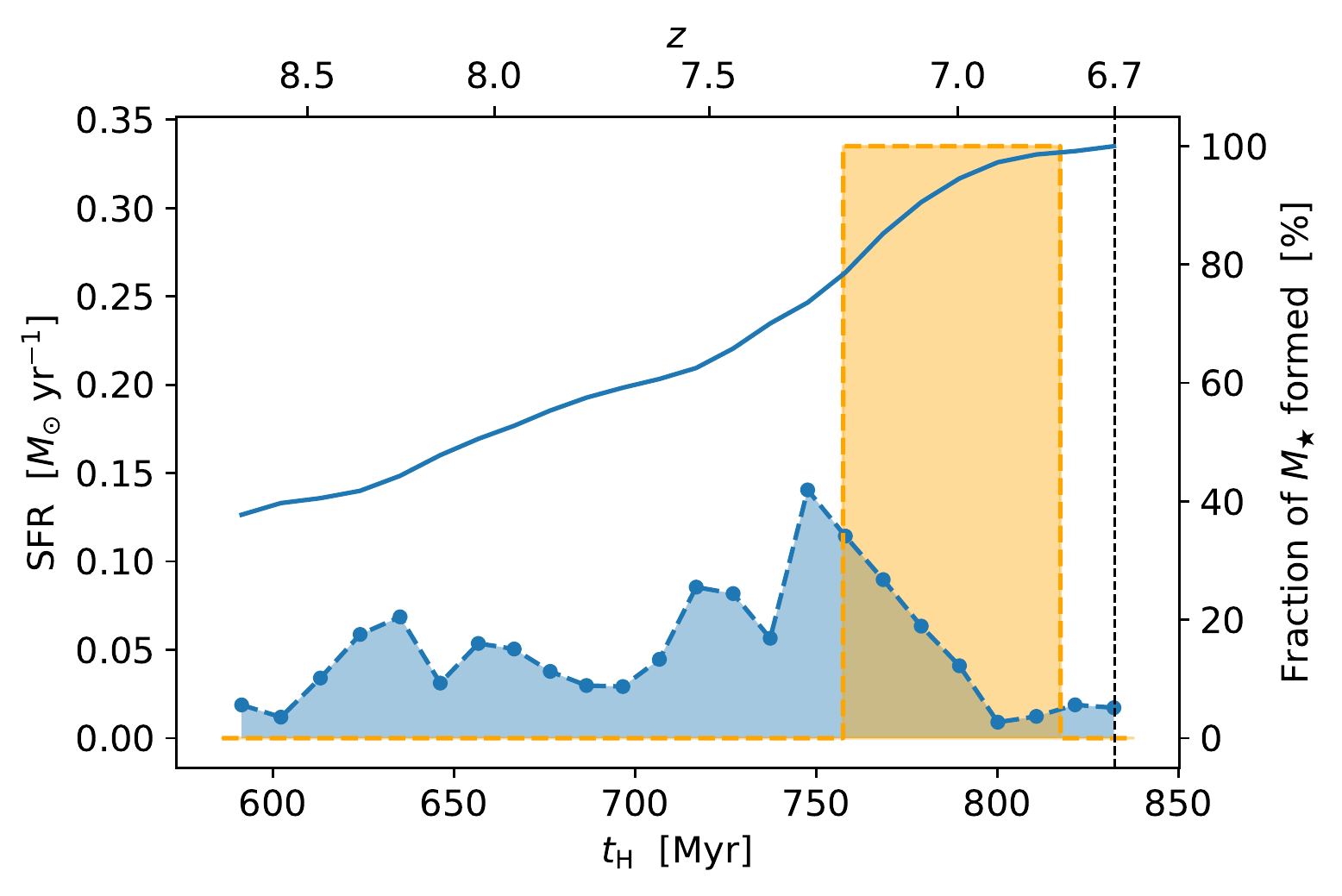}
\end{subfigure}
\begin{subfigure}{0.48\textwidth}
\hspace{0.3cm}
\includegraphics[scale=0.6]{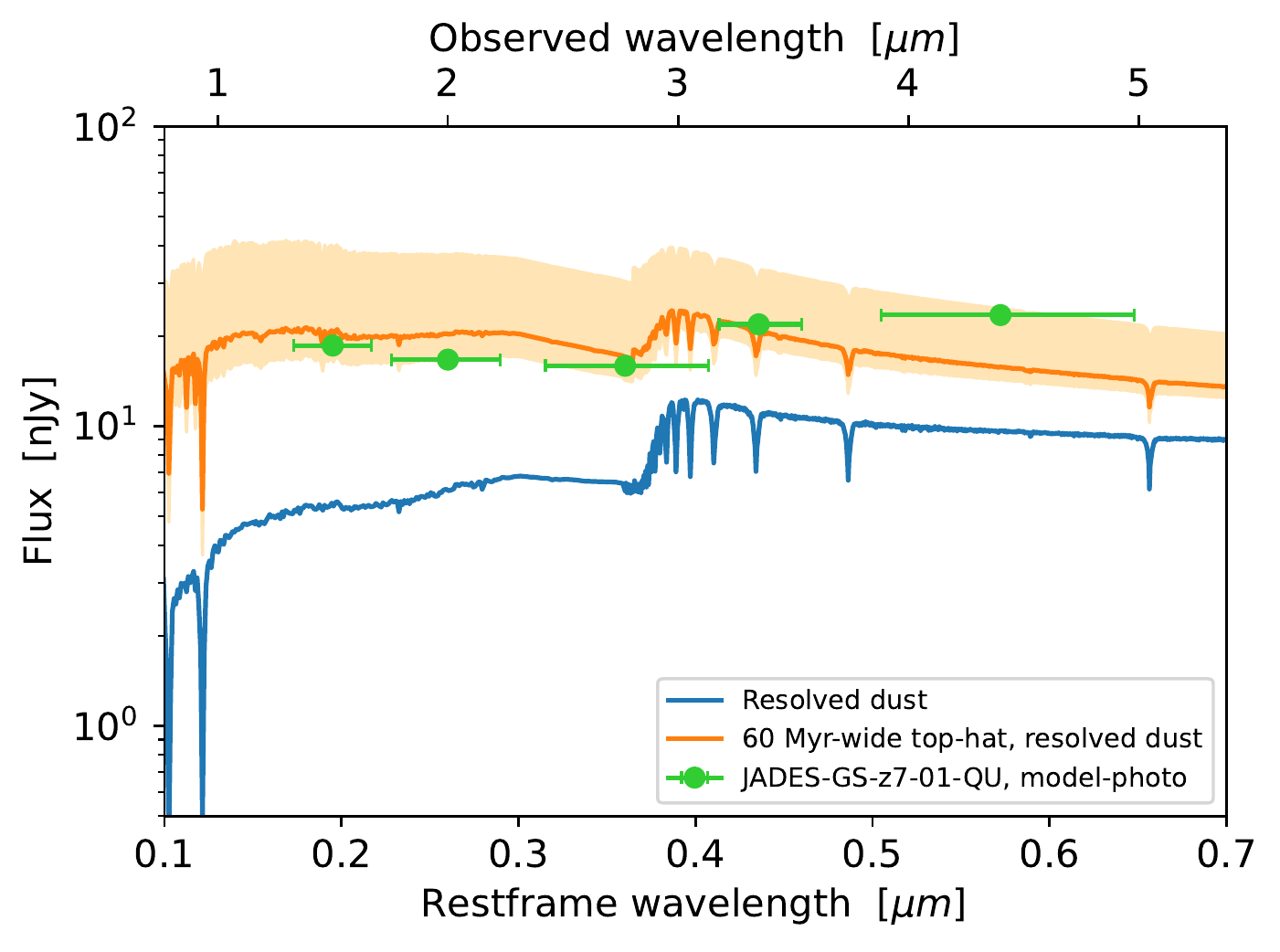}
\end{subfigure}
\caption{Original and modified SFH of the \VELA \ galaxy \scshape{V5} \normalfont at $z>z_{\text{MQ}}=6.6$ (left panel) and corresponding SEDs compared to JADES-GS-z7-01-QU (right panel). Shown are: (a) Original SFH and dust attenuated SED of \scshape{V5} \normalfont (blue), (b) Idealised top-hat SFH model of width $60$ Myr (orange), with corresponding dust attenuated SED (orange), (c) JADES-GS-z7-01-QU photometry (green dots) for filters F150W, F200W, F277W, F335M and F444W. Horizontal bars on the photometry indicate the wavelength range probed by each passband. All SEDs are flux-corrected to account for redshift and stellar-mass offsets between \scshape{V5} \normalfont and JADES-GS-z7-01-QU. The orange shade on the right panel is obtained by varying the width of the top-hat from $10$ to $90$ Myr. The level of burstiness inferred from \VELA \ on short timescales $\sim 40$ Myr is lower than what is observed.}
\label{f_seds}
\end{figure*}
\section{Mock Spectral Energy Distributions Compared to JADES-GS-z7-01-QU}
\label{s_sed_jades}
To bridge the gap with observations, we search for the \VELA \ mini-quenching event that is most similar to JADES-GS-z7-01-QU ($z=7.3$ and $M_{\star} = 5 \times 10^8 \ M_{\odot}$). \VELA \ galaxies at $z\approx 7$ span the mass range $M_{\star} \sim 5.3 \times 10^6 -7.4 \times 10^8 \ M_{\odot}$, yet the few galaxies with $M_{\star} \gtrsim 10^8 \ M_{\odot}$ are far from undergoing mini-quenching. We thus focus on one galaxy (labeled \scshape{V5} \normalfont and highlighted in Fig. \ref{f_dms}) \normalfont and its mini-quenching event at $z=6.7$ and $M_{\star} = 2.1 \times 10^7 \ M_{\odot}$ as the reference galaxy to compare to JADES-GS-z7-01-QU.\par

The account for differences in redshift and stellar mass between \scshape{V5} \normalfont and JADES-GS-z7-01-QU, we apply flux correction factors $d_L(7.3)^2/d_L(6.7)^2$ ($d_L(z)$ being the luminosity distance at redshift $z$) and $M_{\star, \text{V5}}/M_{\star, \text{GS-z7-01-QU}}$ \citep[assuming mass-to-light ratio scaling, see][]{Schombert_2019}. However, incorporating these correction factors we find that the flux density of JADES-GS-z7-01-QU is about a factor of $5$ stronger than \scshape{V5} \normalfont (see Fig. \ref{f_seds}, right panel). Can this discrepancy be resolved by modifying metallicities and ages of the simple stellar populations entering the mock SED calculation?\par 

\subsection{Stellar Metallicity}
Stellar populations in \scshape{V5} \normalfont are only moderately more metal-rich ($Z_{\star}\sim 0.04 \ Z_{\odot}$ on average) than deduced for JADES-GS-z7-01-QU, $Z_{\star}\sim 0.01 \ Z_{\odot}$. Reducing $Z_{\star}$ artificially for some stellar populations does not modify SEDs by more than $\sim 0.1$ dex, in accordance with \cite{Gelli_2023} (at most $\sim 0.2$ dex for a reduction of $Z_{\star}$ by $2$ orders of magnitude from $Z_{\star}\sim Z_{\odot}$ to $Z_{\star}\sim 0.01 \ Z_{\odot}$), hence we do not explore this path further.

\subsection{Ages of Stellar Populations}
The SFH of \scshape{V5} \normalfont (see Fig. \ref{f_seds}, left panel) reveals that the simulated galaxy underwent an extended burst of star formation at $z\approx 7.3$, followed by a gradual suppression of its SFR. At the epoch of observation, $z=6.7$, the average age of stellar populations is thus $\sim 230$ Myr. While the shape of the SED (see Fig. \ref{f_seds}, right panel) agrees moderately well with JADES-GS-z7-01-QU photometry, the overall normalization does not.\par 

Using \scshape{BAGPIPES}, \normalfont \cite{Looser_2023} infer \normalfont a top-hat-like SFH and the time elapsed between mini-quenching and the epoch of observation inferred by four different full spectral fitting codes is $\Delta t_{\text{quench}} = 10-40$ Myr. We thus evaluate the SED of an idealised top-hat SFH. We adopt $\Delta t_{\text{quench}} = 15$ Myr and vary the top-hat width between $10$ Myr and $90$ Myr. The corresponding top-hat heights are $\text{SFR}=2.02 \ M_{\odot}\text{yr}^{-1}$ and $\text{SFR}=0.22 \ M_{\odot}\text{yr}^{-1}$, respectively, and the resulting SEDs are shown as the shaded area in Fig. \ref{f_seds} (right panel). The top-hat that gives rise to an SED closest to JADES-GS-z7-01-QU photometry is of width $60$ Myr (cf. Fig. \ref{f_seds}, left panel). The fluxes now reach $\sim 5$ times higher in the UV and $\sim 2$ times higher in the red part of the spectrum (observed frame), in good agreement with JADES-GS-z7-01-QU. Since the mock SED flux for wavelengths around the red filter F444W is too low compared to JADES-GS-z7-01-QU photometry, the agreement found for the SED shape is only modest. 

\subsection{Caveats}
\label{ss_caveats}
When artificially modifying SFHs and comparing to observations, it is important to mention two caveats. First, \scshape{V5} \normalfont is mini-quenching at $z=6.7$ and $M_{\star} = 2.1 \times 10^7 \ M_{\odot}$ as opposed to the inferred epoch of observation $z=7.3$ and estimated mass $M_{\star} \sim 5 \times 10^8 \ M_{\odot}$ found for JADES-GS-z7-01-QU. The mass-to-light scaling that we adopt might be an invalid assumption. However, we have repeated the SED analysis for several mini-quenching events in \VELA \ across $z=6.5-7.5$. For \IBox, we have likewise calculated SEDs of MQGs at $z=7$, which is the closest redshift at which a resolved dust attenuated SED modelling can be performed\footnote{The majority of \TNG \ snapshots only have a subset of particle fields available.}, following \cite{Nelson_2018}. In all cases, we come to the same conclusion: The SEDs can only be reconciled with JADES-GS-z7-01-QU photometry when artificially modifying ages of stellar populations.\par 

Secondly, there are different recipes for dust attenuation modelling, and while an extensive comparison is beyond the scope \citep[cf.][]{Nelson_2018, Vogelsberger_2020, Shen_2020}, we find that differences between e.g. adopting HI vs H column densities only lead to minute effects on the resulting mock SEDs. However, since JADES-GS-z7-01-QU photometry suggests a redder spectrum than we infer for the top-hat SFHs, we speculate that JADES-GS-z7-01-QU is more dust-obscured than \scshape{V5} \normalfont and/or has some older populations than the top-hat.\par 

\subsection{Discussion}
\label{ss_discussion}
We speculate that \VELA \ and \TNG \ galaxies at high redshift are not bursty enough on small timescales $\sim 40$ Myr to give rise to the high fluxes observed for JADES-GS-z7-01-QU. Only when allowing a maximum burst of star formation (top-hat SFH), matching $\Delta t_{\text{quench}} = 15$ Myr with the observed one, can the normalization of SEDs be reconciled. While the exact value observed in the red filter F444W is hard to reproduce in mocks, the overall SED shape inferred for a top-hat SFH is in good agreement with JADES-GS-z7-01-QU, including Balmer absorption lines, lack of emission lines, and UV continuum. There is a possibility that JADES-GS-z7-01-QU is in fact an obscured AGN, which would not only provide a mechanism for abrupt quenching but possibly explain the flux discrepancy in the F444W filter. The abundance of (dust-obscured) AGN at high redshift of $z>5$ might be an order of magnitude higher than expected from extrapolating quasar UV luminosity functions \citep{Matthee_2023, Larson_2023, Ubler_2023, Endsley_2023}. However, unlike JADES-GS-z7-01-QU many high-redshift AGN appear to have a strong Balmer break \citep[e.g.][]{Kocevski_2023}.\par 

Which sub-grid models governing galaxy formation could give rise to higher burstiness on small timescales ($\sim 40$ Myr) to match the observations? While for \TNG \ galaxies, resolution effects \citep[e.g. $M_{\star}$ of low-mass galaxies typically increases with resolution, see][]{Pillepich_2018} might also be at play, \TNG \ has been shown to poorly predict some observables at high redshift, e.g. the scatter around the MS at the high-mass end (cf. Sec. \ref{ss_scatter_ms}) or the abundance of dust-obscured, far-infrared galaxies and thus the obscured cosmic star formation rate density \citep{Shen_2022}.\par 

One possibility is that better modelling at high redshift resolves the discrepancies with observations found here. The fact that \cite{Gelli_2023} also succeeds in reproducing the SED of JADES-GS-z7-01-QU when artificially modifying SFHs points in that direction. They analyse the \mbox{\scshape{SERRA}} \normalfont suite of high-resolution zoom-in simulations that includes on-the-fly radiative transfer and a non-equilibrium chemical network. Even though these prescriptions are more suitably in the epoch of reionization \citep[see][]{Pallottini_2022} than a spatially uniform UV background \citep{Haardt_2012}, the levels of burstiness they infer on small timescales $\sim 40$ Myr are still too low.

\section{Conclusions}
\label{s_conclusions}
Bursty star formation at high redshift gives rise to (likely only) temporarily quenched, or mini-quenched galaxies in the mass range $M_{\star} = 10^7-10^9 \ M_{\odot}$. With the advent of JWST and the first observations of such galaxies \citep{Looser_2023, Strait_2023}, it is critical to gain a thorough understanding of the physical mechanisms and the timescales involved. Combining insights and leveraging periodic box simulations, zoom-in simulations and empirical models is an important first step in understanding the regulation of star formation in high redshift galaxies.\par

\textit{Methods:} We employ four galaxy formation models, a periodic box simulation (\TNG), two zoom-in simulations (\FL\ and \VELA) and an empirical halo model (EHM) to investigate the properties of high-redshift (mini-)quenched galaxies. We adopt an aperture of twice the stellar half mass radius $2\times R_{\mathrm{star,h}}$ (and $0.15 \ R_{\text{vir}}$ for \FL), an averaging timescale of $\Delta t = 10$ Myr ($\Delta t = t_{\text{dyn}}$ for EHM) and a selection threshold of $0.75$ dex below the star-forming MS.\par

\textit{Abundance:} We find that the abundance of quenched galaxies at high redshift inferred from \TNG, \VELA \ and EHM is largely consistent with each other, implying that this galaxy population is rare. The quenched galaxy population first appears below $z\approx 8$, after which their fraction increases with cosmic time, from $\sim 0.5-1.0$\% at $z=7$ to $\sim 2-4$\% at $z=4$ in the mass range $M_{\star} = 10^7-10^9 \ M_{\odot}$. The corresponding comoving number densities read $10^{-5}$ Mpc$^{-3}$ at $z=7$ and $10^{-3}$ Mpc$^{-3}$ at $z=4$. The number of quenched galaxies decreases monotonically with increasing stellar mass $M_{\star}$, a dependence that is stronger in models which probe a smaller dynamic range (smaller box sizes) such as \SBox. Quenched galaxy fractions in \TNG \ and EHM are consistent with SFR duty cycle estimates ($f_{\mathrm{duty}}\sim 98.6-99.9\%$ at $z=7$) inferred for \FL \ and \VELA \ galaxies.\par

\textit{Duration:} For MQGs, SFHs rapidly change before, during and after the mini-quenching event, consistent with the idea that mini-quenching results from bursty star formation. The distribution of mini-quenching timescales (defined as the accumulated time a MQG spends $\geq 0.75$ dex below the MS) averaged across $z=4-8$ in \FL \ and \VELA \ peaks around $\tau_{\mathrm{MQ}} \sim 20-40$ Myr. While the upper tail of the distribution is highly sensitive to the threshold adopted, only one simulated galaxy stays quiescent for an extended period of time (\scshape{V28} \normalfont for $\sim 220$ Myr at redshifts $z\approx 4.9-5.7$). This mini-quenching timescale is close to the local free-fall timescale $t_{\mathrm{ff}}$ of the \textit{inner halo}.\par

\textit{Cause:} In EHM, quenching is by construction caused by a lack of gas inflow, which itself is tied to dark matter accretion rates. In \TNG, \FL \ and \VELA, we in addition find that the periodic injection of energy and momentum into the circum- and intergalactic medium via stellar feedback in the context of bursty star formation contributes to the regulation of star formation and thus the phenomenon of mini-quenching. However, by following the main progenitor branch and main descendant branch of quenched galaxies in \IBox \ and \SBox, we show that many quenched galaxies ($\sim 27$\% at $z=7$) are gravitationally interacting with other galaxies, and even when not fully merging are tidally disrupted.\par

\textit{Consistency with Observations:} Simulated SEDs can only be reconciled with JADES-GS-z7-01-QU photometry in both \TNG \ and \VELA \ when artificially modifying ages of simulated stellar populations. In particular, a top-hat SFH of width $60$ Myr shows best agreement with JADES-GS-z7-01-QU, consistent with observationally inferred SFHs. While simulated SED shapes agree moderately well including Balmer absorption lines, the flux density in the red F444W filter is lower in the top-hat SEDs than observed by a factor of $\sim 1.5$. This is likely caused by higher levels of dust obscuration for JADES-GS-z7-01-QU compared to simulated galaxies, though some older populations would be needed for even better agreement, disallowed by the top-hat. Alternatively, JADES-GS-z7-01-QU could be an obscured AGN.\par

\textit{Outlook:} The fact that we need to artificially modify ages of stellar populations to find agreement with the observed SED lets us conclude that sub-grid models governing galaxy formation at high redshift have likely to be adapted, including in the higher-mass regime of $M_{\star} = 10^8-10^9 \ M_{\odot}$ in which star formation is expected to transition from bursty to stable. On-the-fly radiative transfer and a non-equilibrium chemical network \citep{Gelli_2023} adopted for \mbox{\scshape{SERRA}} \normalfont simulations is not enough to remedy the discrepancies. In the context of bursty star formation, an improved understanding is needed of how much (stochastic) power exists on the temporal scales probed by observations. MQGs can thus be a useful probe for sub-grid models, and will help close the gap between observations and theoretical models. Extending the concept of MQGs to lower redshift and studying the transition from bursty to steady star formation at $z=1-3$ will be necessary to interpret the upcoming wealth of measurements on the low-mass quiescent population driven by deep JWST data.

\section{Acknowledgements}
It is a pleasure to thank Debora \v{S}ija\v{c}ki for enriching conversations. We thank Takumi Tanaka, who inspired us to use ``mini-quenching'' for describing short-term quenching. We are grateful to our anonymous referee for providing valuable feedback that improved the quality of our manuscript. TD acknowledges support from the Isaac Newton Studentship and the Science and Technology Facilities Council (STFC) under grant number ST/V50659X/1. AF is supported by the Royal Society University Research Fellowship. AD, SL and OG were partly supported by the Israel Science Foundation grant 861/20. OG is supported by a Milner Fellowship. T.J.L acknowledges support by STFC and ERC Advanced Grant 695671 ``QUENCH''.

\section{Data Availability}
\label{s_data_availability}
The \TNG \ simulation snapshots are publicly accessible at \href{https://www.tng-project.org/}{https://www.tng-project.org/}. SFHs of \FL \ are available at \href{http://odin.ft.uam.es/FirstLight/index.html}{http://odin.ft.uam.es/FirstLight/index.html}. Post-processing scripts and EHM data are made available upon reasonable request.

\bibliographystyle{mnras}
\bibliography{refs}

\label{lastpage}
\end{document}